%% file: PaidDataFinal.tex
\documentclass[12pt,a4paper]{scrartcl}
\usepackage{amsmath,amssymb,subfigure,graphicx}
\usepackage{etoolbox}             

\usepackage{dsfont}   
\usepackage{authblk}   
\usepackage{natbib}   
\usepackage{times}   

\let\originalleft\left
\let\originalright\right
\renewcommand{\left}{\mathopen{}\mathclose\bgroup\originalleft}
\renewcommand{\right}{\aftergroup\egroup\originalright}

\newtheorem{theorem}{Theorem}[section]
\newtheorem{lemma}[theorem]{Lemma}
\newtheorem{proposition}[theorem]{Proposition}
\newtheorem{corollary}[theorem]{Corollary}

\newenvironment{proof}[1][Proof]{\begin{trivlist}
\item[\hskip \labelsep {\bfseries #1}]}{\end{trivlist}}

\newenvironment{remark}[1][Remark]{\begin{trivlist}
\item[\hskip \labelsep {\bfseries #1}]}{\end{trivlist}}
 
\DeclareMathOperator{\ISE}{ISE}
\DeclareMathOperator{\err}{err}
\DeclareMathOperator{\MSE}{MSE}

\begin{document}
\title{Continuous chain-ladder with paid data}  
\author[1]{Stephan M.~Bischofberger\footnote{Corresponding author: Stephan M.~Bischofberger, e-mail: stephan.bischofberger@cass.city.ac.uk, address: Cass Business School, 106 Bunhill Row, London, EC1Y 8TZ, United Kingdom.}}
\author[2]{Munir Hiabu}
\author[1]{Alex Isakson}
\affil[1]{Cass Business School, City, University of London, United Kingdom}
\affil[2]{School of Mathematics and Statistics, University of Sydney, Australia}
\maketitle

\AtBeginEnvironment{tabular}{\footnotesize} 

\begin{abstract}
We introduce a continuous-time framework for the prediction of outstanding liabilities, in which chain-ladder development factors arise as a histogram estimator of a cost-weighted hazard function running in reversed development time. We use this formulation to show that under our assumptions on the individual data chain-ladder is consistent. Consistency is understood in the sense that both the number of observed claims grows to infinity and the level of aggregation tends to zero.
We propose alternatives to chain-ladder development factors by replacing the histogram estimator with kernel smoothers and by estimating a cost-weighted density instead of a cost-weighted hazard. Finally, we provide a real-data example and a simulation study confirming the strengths of the proposed alternatives. 
\end{abstract}


\section{Introduction} \label{sec:introduction}
The classical run-off triangle used for the prediction of outstanding liabilities  can be explained as a two-way ANOVA arrangement, where data  is organized on a two-dimensional plane of (cohort, age) with cohort being the accident date or underwriting date of a claim, and age the time from that date to a payment. 
Developed at least in the beginning of the last century,  the chain-ladder method is still the industry standard for estimating the future cost of outstanding liabilities
from these run-off triangles. 
However, as a deterministic algorithm, chain-ladder does not specify the assumptions that it is based on, nor the uncertainty of the estimation.

\begin{figure}[h!] 
\centering
\subfigure[][Individual payments  in aggregated cells.]{
\includegraphics[width=0.45\textwidth]{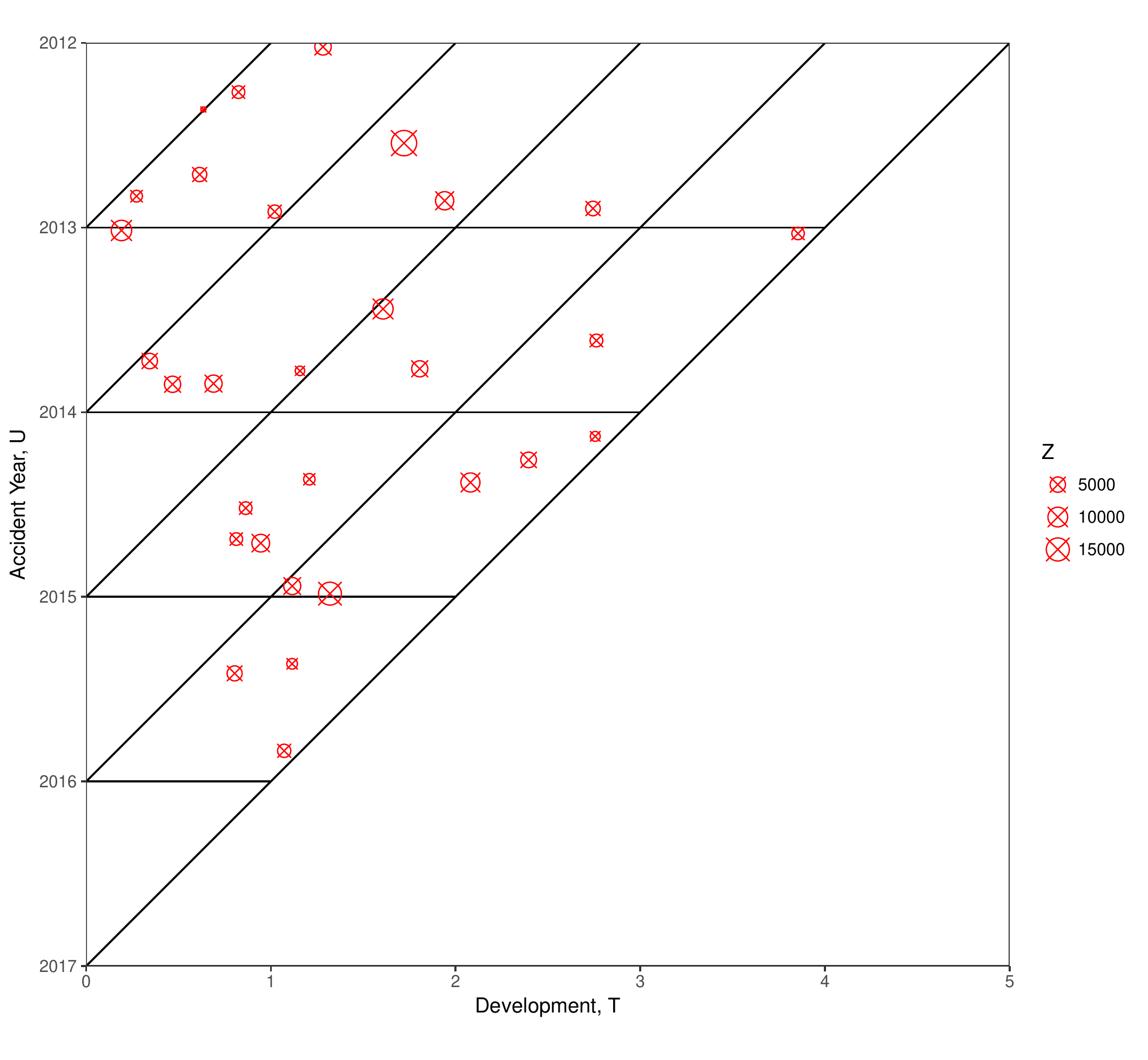}
}
\subfigure[][Individual payments with no grid.]{
\includegraphics[width=0.45\textwidth]{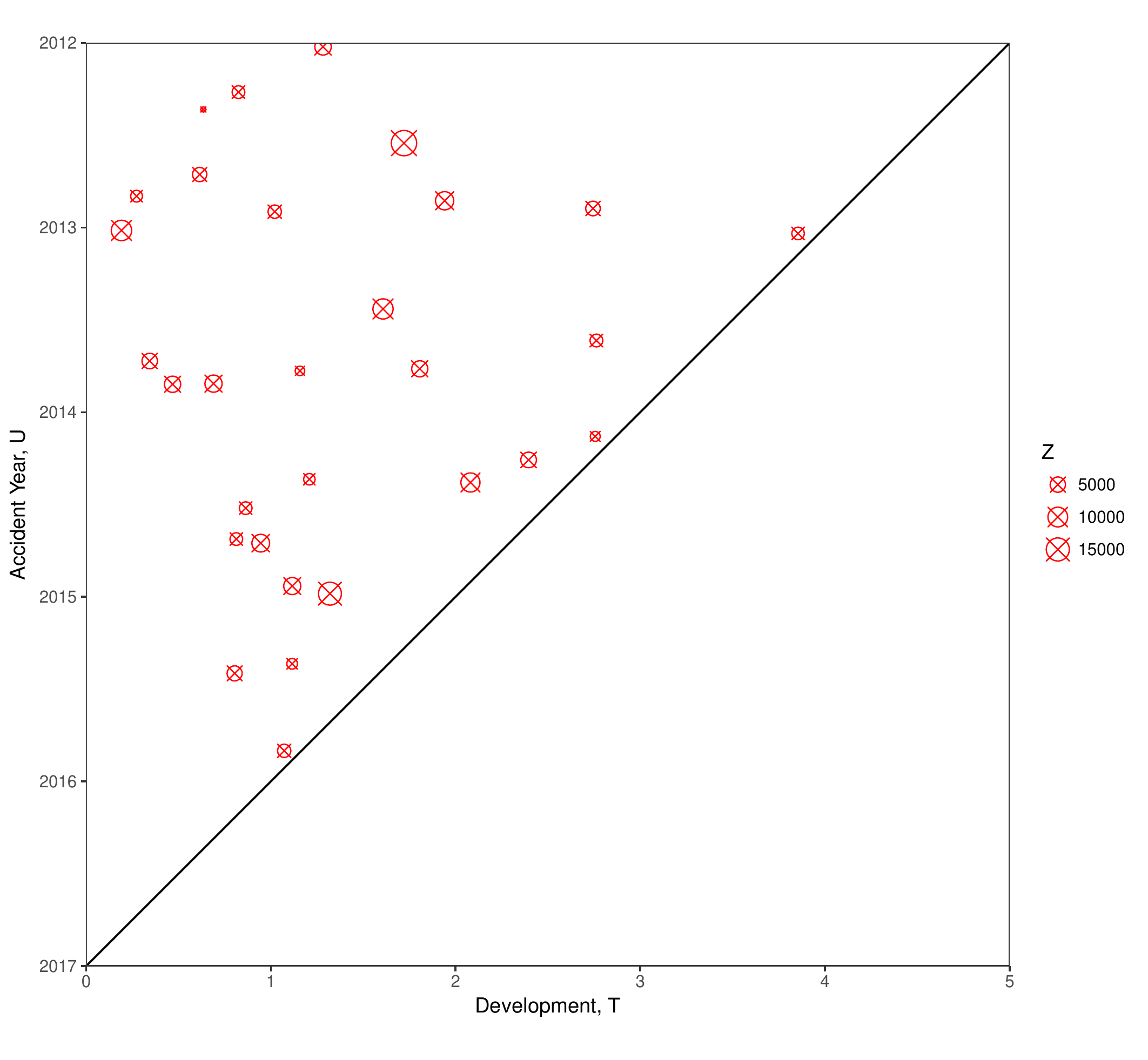}
}
\caption{Payments are independent and have three features: accident date $U_i$, development delay $T_i$, and claim severity $Z_i$.}
\label{fig:data0}
\end{figure}

Stochastic models around the chain-ladder method are the Mack Model \citep{Mack:93} and multiplicative models in 
 \cite{Kremer:82,Verrall:91,Renshaw:Verrall:98} and \cite{Kuang:etal:09} among many others. A comprehensive review is given in \cite{England:Verrall:02}. 
The drawback of these papers is that they do not discuss how the data arises as aggregation from individual data. This is needed when one wants to understand the underlying assumptions of the model. \cite{Taylor:86} coined the term macro-models to describe these previous models and defined models that begin on an individual level as micro-models. 
Macro models have assumptions which are hard to justify once a data generating process with individual payments is considered.
The assumptions of the most widely used Mack model 
can hardly be justified if one considers that the cells in the classical run-off triangle are 
aggregations of individual payments. Under Mack's assumptions, not a single future payment can be independent of the past. This is because the conditional
expectation of the next cell within a row of the run-off  triangle is a multiple of all previous observations in the same row.
The other big class of models is those of \cite{Kremer:82,Verrall:91,Renshaw:Verrall:98} and \cite{Kuang:etal:09}. They  assume that the expected claim amount in one cell  is the product of a row factor and a column factor --- representing underwriting/accident date  and payment delay, respectively.
This multiplicative structure implies that there is no interaction effect between rows and columns working on the expected claim amount.
In \cite{Hiabu:17} it has been shown that this non-interaction assumption generally does not hold because  the cells in the  run-off triangle are aggregated as parallelograms as illustrated in Figure \ref{fig:data0}.
These parallelograms will generally introduce interdependencies, which violate the multiplicative structure assumption leading to an interaction effect. Hence, assuming a multiplicative structure produces a bias that grows with the level of aggregation. Therefore, as done in this paper, consistency of payment  predictions can only hold in a continuous framework where the level of aggregation is understood to converge to zero with increasing number of observations. 

Recent literature  connects the chain-ladder method and its data to counting process theory in survival analysis.
\cite{Hiabu:etal:16} introduced a statistical model including the data generating process 
which is built on the continuous model of \cite{Martinez:etal:13}.
The sampling technique of the chain-ladder method is different from other sampling techniques used in classical  (bio-)statistical literature.
Individuals or policies are only followed if a failure, i.e., a claim occurs.
This has the advantage  that less data is required than in classical survival data. 
Truncation occurs when $\text{cohort}$ plus $\text{age}$ is greater than the date of data collection.
However, \cite{Martinez:etal:13} and
\cite{Hiabu:etal:16} only considered 
 claim counts, and ignored its associated payments. 

In this paper, we introduce a micro-model in continuous time in which chain-ladder development factors, applied on a paid triangle, are a histogram estimator of a cost-weighted hazard function running in reversed development time. 
We establish new assumptions under which consistency of the development factors is achieved.
Consistency is understood in the sense that both the number of observed claims grows to infinity and the level of aggregation tends to zero. 
Finally, we improve on chain-ladder estimation by replacing the histogram estimator with kernel smoothers and by estimating a cost-weighted density instead of a cost-weighted hazard.

There is also a growing literature on micro-models for estimating outstanding liabilities in non-life insurance that is not based on the chain-ladder idea.
\cite{Arjas:89} and \cite{Norberg:93} formulated models in a classical bio-statistical setup with a non-homogeneous marked Poisson process.
A  strong case study in this setting has been developed in \cite{Antonio:Plat:14} and the models have been further studied in \cite{Huang:etal:15} and \cite{Huang:etal:16}.
These models are more complex than chain-ladder models. They model each delay component in the claims process separately and require full inference on the marked point process, for instance the distribution of the mark/cost. They also require additional information about the exposure, i.e.,  information about the number of policies underwritten. 
The assumptions of this paper can be used to decide whether this additional complexity is beneficial noting that additional complexity introduces bias and is only advisable if a significantly better fit can be obtained.
Additional complexity might also be necessary if claims with different accident dates, e.g., due to calendar time effects, are not independent, as investigated in  
\cite{Shi:etal:12, Merz:etal:13, Lee:etal:15,  Badescu:etal:16, Avanzi:etal:16, Lee:etal:17} and \cite{Crevecoeur:etal:19}.
If more complexity is justified, 
the estimator presented in this paper can be used as a building block in those and other more complex models. 
However, this is beyond the scope of this present paper.

This paper is structured as follows. Section 2 describes the mathematical model and Section 3 links chain-ladder development factors to that framework by identifying them as histogram estimator of a hazard function. Section 4  proposes improvements on chain-ladder development factors by replacing the histogram with kernel smoothers and by estimating a density function instead of a hazard function. 
We provide a data application and simulation studies in Sections 5 and 6. All proofs can be found in the appendix.

\section{Mathematical framework} \label{sec:math:framework}
We start by putting the unique sampling scheme of chain-ladder into a micro-structure framework.
We observe counting processes $(N_i(t))_{t\in [0,\mathcal T]}$, $\mathcal T>0$, for claims $i=1, \dots n$  and call $t$ development time.
Each counting process starts with value zero at the underwriting date underlying its claim. It  jumps, with jump-size one, whenever a payment is made. Additionally to every jump, we observe a mark indicating the size of the payment made.
The number of counting processes, $n$, varies  over calendar-time: 
We follow retrospectively only those claims for which at least one payment has been observed, i.e.,  we do not follow every claim in the policy book.
In this paper, we make the following assumptions. 
\begin{enumerate}
    \item[{[M1]}] All claims are independent.
    \item[{[M2]}] Every claim consists of only one payment.
\end{enumerate}
Assumptions [M1] and [M2] are rather strong but are made to simplify the mathematical derivations yielding a first and clean step towards a better understanding of chain-ladder on a micro-structure level. Possible ways to relax these assumptions are weak dependency (instead of [M1])
and a Markov process structure where every jump triggers a new state (instead of [M2]). This, however, is beyond the scope of the present paper.

The jump-time in development direction corresponding to the payment for claim $i$ is denoted by $T_i$. 
Thus, we get
\[
N_i(t)=I(t\geq T_{i}),
\]
where $I(\cdot)$ denotes the indicator function.
As pointed out in \cite{Hiabu:etal:16}, 
statistical inference on the counting process $N_i$ is not directly feasible.
We only follow a claim once we have observed at least one payment. Therefore, by design it holds
\[
 T_{i}\leq today -U_i,
\]
where $U_i$ is the underwriting date or accident date of claim $i$. 
Hence, by not following every policy, we are exposed to a
right-truncation problem instead of a right-censoring problem.
In the sequel, for notational convenience, we parameterize the dates such that $today = \mathcal T$ which yields $T_{i}\leq \mathcal T -U_i$.

A solution to the right-truncation problem is to reverse the time of the counting process leading to a tractable left-truncation problem \citep{Ware:DeMets:76}.
To this end we consider the counting processes
\[
N_i^R(t)= I(t\geq  T^R_{i}), \quad  T^R_{i}=\mathcal T- T_{i},
\]
each with respect to the filtration
\begin{equation*}
\mathcal F^R_{it}=\sigma \left( \bigg\{T^R_i \leq s:\ s\leq t\bigg\} \cup \bigg\{    U_i \leq s:\  s\leq t\bigg\} \cup \mathcal N\right), 
\end{equation*}
satisfying the \textit{usual conditions} \citep[p.~60]{Andersen:etal:93}, and where $\mathcal N$ is the set of all zero probability events.
It is well known \citep[Theorem II.4.2]{Andersen:etal:93},
that the intensity process of $N_i^R$ is
\[
\lambda_i^R(t)=\lim_{h \downarrow 0} h^{-1}E\left[  N^R_i\left\{(t+h)-\right\}- N^R_i(t-)|\ \mathcal F^R_{it-}\right]
=\alpha^R(t)Y_i^R(t),
\]
where 
\begin{align*}
\alpha^R(t)&=\lim_{h \downarrow 0} h^{-1}{P}\left(T_i^R\in[t,t+h)|\ Y_i^R(t)=1\right), \\
Y_i^R(t)&=I(U_i< t\leq T^R_i),
\end{align*}
which is a product of a deterministic function and a predictable function.
This structure is called Aalen's multiplicative intensity model 
\citep{Aalen:78}, and enables nonparametric estimation and inference 
on the deterministic factor $\alpha^R(t)$, which is done in  \cite{Hiabu:17}.

Let $Z_i$ denote the payment size of claim $i$
and consider the process $\widetilde N^R_i(t)=Z_i N_i^R(t)$. 
Ignoring for now the necessary regularity conditions, 
it is straight forward to see that 
\begin{align*}
\widetilde \lambda^R_i(t)&=\lim_{h \downarrow 0} h^{-1}E\left[  \widetilde N^R_i\left\{(t+h)-\right\}- \widetilde N^R_i(t-)|\ \mathcal F^R_{it-}\right]=\mathcal R_n(t) \widetilde \alpha^R_\ast(t)\widetilde Y_i^R(t),\\
\widetilde \alpha_\ast^R(t)&=
\frac{ \lim_{h \downarrow 0}  E[Z_1 | T_1\in [t,t+h)]}{ E\left[Z_1 | \ Y^R_1(t)=1\right]} \alpha^R(t),\\
\widetilde Y_i^R(t)&=\frac{\sum_j Z_j Y^R_j(t)}{ \sum_j Y^R_j(t)}Y^R_i(t),\\
\mathcal R_n(t)&= \frac{E{\left[Z_1 | \ Y^R_1(t)=1\right]} }{\sum_j Z_j Y^R_j(t)/ \sum_j Y^R_j(t)}, 
\end{align*}
which asymptotically satisfies Aalen's multiplicative intensity structure, if $\mathcal R_n(t)$ converges to 1. This convergence will be verified below and it is sufficient to apply the well developed techniques for counting processes, which we do in this paper.
In the next section we will show that chain-ladder development factors (which are defined for instance in \cite{Taylor:86}) are a nonparametric histogram estimator of $1+\widetilde \alpha_\ast^R(t)$.

When the goal is to predict outstanding liabilities, one is interested in the untruncated versions of the truncated observations. We will indicate these variables by suppressing the subscript, i.e., three-dimensional random variable $(T,U,Z)$ has the same distribution as
$(T_i,U_i,Z_i)$, for every $i=1,\dots,n$, if conditioned on the event $\{T\leq \mathcal T - U \}$.
We make the following assumptions on the untruncated objects.
\begin{enumerate}
\item[{[M3]}] The random variables $T$ and $U$ have, respectively, strictly positive continuous density functions $f_T$ with support $[0,\mathcal T]$ and $f_U$  with support $[0,\mathcal U]$, $\mathcal U \leq \mathcal T$, each  with respect to the Lebesgue measure. Moreover, the continuous joint density $g$ of $(T,U,Z)$ with respect to the Lebesgue measure exists and $E[|Z|]<\infty$ 
\item[{[CLM1]}]
The random variables $T$ and $U$ are independent.

\item[{[CLM2]}]
There exist functions $m_1, m_2$ such that $E[Z | T,U]=m_1(T)m_2(U)$.

\end{enumerate}
Assumption {[M3]} ensures that the intensity $\widetilde \lambda_i^R$ is well defined.
Note that [CLM1] is a statement about the untruncated objects and does not imply that $U_i$ and $T^R_i$ are independent, noting that $U_i \leq T^R_i$. The second part of 
Assumption [CLM2] means that there is no interaction effect between development time and underwriting date on the claim amount. \\

To align the density of $T$ with our setting, we define the cost-weighted density of $T$ as 
\[
\widetilde f_T(t)=\frac{E\left[Z| \ T=t \right]} { E[Z]} f_T(t).
\]
Conditional expectations to point events with probability zero here and below are well defined through the continuous density $f_T$. 
Analogously, we define the cost-weighted density in reversed time as $\widetilde f_T^R(t) = \widetilde f_T(\mathcal{T}-t)$ and $f_T^R(t) = f_T(\mathcal{T}-t)$. 
Moreover, the underlying hazard rate in reversed time can be derived from the above definition as
\[
\widetilde \alpha^R(t)= \frac{\widetilde f^R_T(t)}{\widetilde S^R_T(t)} = \frac{E\left[Z| \ T^R=t \right]} { E[Z| T^R\geq t]} \frac{f_T^R(t)}{\int_{t}^{\mathcal T}f_T^R(s)\mathrm ds}.
\]

\begin{proposition}\label{prop11} 
Given [M1]--[M3], for $n\to \infty$, it holds
\begin{equation}\label{prop:CLM:assumption}
 \sup_{t\in[0,\mathcal{T}]}  \mathcal R_n(t) \rightarrow 1.
 \end{equation}
 If additionally Assumptions [CLM1] and [CLM2] hold, then 
 \begin{equation}
 \label{prop:CLM:assumption2}
 \widetilde \alpha_\ast^R(t) = \widetilde \alpha^R(t).
\end{equation}
\end{proposition}
\begin{proof}
See Appendix \ref{proof:prop1}.
\end{proof}

Proposition \ref{prop11} explains why we gave the last two assumptions a  `CLM'-prefix. 
The convergence in \eqref{prop:CLM:assumption} ensures that
Aalen's multiplicative intensity model is approximately satisfied and chain-ladder development factors are histogram estimates of $1+\widetilde \alpha_\ast^R$. But it is only via equation \eqref{prop:CLM:assumption2} that chain-ladder development factors  also approximate $1+\widetilde \alpha^R$. For the latter to be true,  we  assume  [CLM1] and [CLM2]. Under these two additional conditions,  the chain-ladder algorithm  predicts the right object and leads to a  sensible quantification of the outstanding liabilities.

We close this section with some further remarks.
\begin{remark}[Remark 1](Exposure).
 As in traditional chain-ladder and in contrast to  classical survival data, because all failures (claims) are observed, there is no additional information needed about the number of individuals under risk (i.e.\ exposure in form of the number of underwritten policies) in order to estimate future claim amounts. The unique sampling leads to a right-truncation which is solved by reversing the time.
This is different to the approaches described in \cite{Arjas:89} and \cite{Norberg:93}.
\end{remark}

\begin{remark}[Remark 2](Cost-weighted density).
The density $\widetilde f_T$ is the continuous analogue to the column parameter of chain-ladder, which is often called $\beta$ and which is considered in  \cite{Kremer:82,Verrall:91,Renshaw:Verrall:98} and \cite{Kuang:etal:09}. Moreover, integrating to one,  $\widetilde f_T$ is indeed a density function. 
\end{remark}

\begin{remark}[Remark 3](Predicting outstanding liabilities). \label{remark:total:outstanding}
An estimator of  the cost-weighted hazard, $\widetilde \alpha^R(t)$,
in conjunction with a chain-ladder algorithm can be used to predict outstanding liabilities. Alternatively, as proposed in this paper,  one can employ estimators of the cost-weighted densities
$\widetilde f_T, \widetilde f_U$.
If the maximum development time of a claim is $\mathcal T$, then
the expected outstanding liabilities for claims underwritten in $[0,\mathcal U]$, $\mathcal U\leq \mathcal T$, is given as 
\begin{equation}\label{rn}
r_n = \frac{\int_0^{\mathcal U} \int_{1-u}^{\mathcal T}  \widetilde f_{T,U}(t,u) \mathrm dt \mathrm du }{ \int_0^{\mathcal U} \int_0^{\mathcal T -u}  \widetilde f_{T,U}(t,u) \mathrm dt \mathrm du } \sum_{i=1}^n Z_i, 
\end{equation}
where $\widetilde f_{T,U}(t,u)=E[Z]^{-1} E\left[Z| \ T=t , U=u \right]f_{T,U}(t,u)$ is the cost-weighted density of $(T,U)$. 
The total amount of payments until today is given by $\sum_{i=1}^n Z_i$ and the fraction in $r_n$ 
gives the expected ratio between outstanding payments and past payments.
Note that under Assumptions [CLM1] and [CLM2],  the cost-weighted joint density factorizes  into $\widetilde f_{T,U}(t,u)=\widetilde f_T(t) \widetilde f_U(u)$.  
 In Section \ref{sec:local:polynomial} we propose estimators for $ \widetilde f_T$.
Due to symmetry, the component $ \widetilde f_U$ can be estimated by swapping the roles of $T$ and $U$. Outstanding liabilities are estimated by replacing $ \widetilde f_T$ and $ \widetilde f_U$ in $r_n$ with their estimates. 
Developing estimation theory for $r_n$ is rather involved because of the non-trivial integrals and the ratio-structure in \eqref{rn}. 
We only consider a simulation study for the prediction performance of $r_n$ in Section 6.
\end{remark}

\begin{remark}[Remark 4](Assumptions).
While the model is built around the observation of independent claims with one single payment each (Assumptions [M1] and [M2]), allowing for claim clusters and multiple payments per claim is feasible and would only require some further assumptions.

Assumption [CLM1] is analogue to the usual multiplicity assumption found for example in  \cite{Kremer:82,Verrall:91,Renshaw:Verrall:98} and \cite{Kuang:etal:09}. The difference is that [CLM1] refers to claim
counts and not to claim amounts.
However, [CLM1] and [CLM2] together imply the multiplicity of aggregated expected claim amounts as assumed in the literature --- if ignoring the potential bias arising from  aggregation. Chain-ladder development factors can be biased if
the cost-weighed  development delay  with density $\widetilde f_T$ is neither exponentially distributed nor uniformly distributed within each development period \citep{Hiabu:17}. 
\end{remark}

\section{Chain-ladder development factors}\label{sec:CLM}
We now discuss how hazard rates can be estimated in the framework of Section 2. 
In  the setting of Proposition \ref{prop11},   the 
intensity of $\widetilde N_i^R$ at $t$ is asymptotically equal to $\widetilde  \alpha^R(t) \widetilde Y_i^R(t)$. 
We use this fact to construct a least squares criterion to estimate  $\widetilde  \alpha^R(t)$.
Given a smoothing parameter, $h>0$, and a weight function $W_h(\cdot,\cdot)$, we look
for estimators  $\widehat {\widetilde \alpha}^R(t)$ that minimize
\begin{equation}\label{ls:hazard}
  \lim_{\varepsilon \downarrow 0}\sum_{i=1}^n \int \left[\left\{ \frac 1 \varepsilon \int_{s}^{s +\varepsilon} 
  \mathrm{d}\widetilde N^R_i(w) -  \widehat {\widetilde \alpha^R} (t)\widetilde Y^R_i(s) \right\}^2 -\xi(\varepsilon)\right] W_h(s,t) \frac 1 {\widetilde Y_i^R (s)} \, \mathrm {d}s,
\end{equation}
where the expression $1 /{\widetilde Y_i^R (s)}$ is  understood as being zero whenever 
$\widetilde Y_i^R (s)$ is zero. The term
 $\xi(\varepsilon)=\{\varepsilon^{-1} \int_{s}^{s+\varepsilon}    \mathrm{d}\widetilde N^R_i(w)\}^{-2}$  is a vertical shift  subtracted to make the integral well-defined.
Since $\xi(\varepsilon)$ does not depend on $\widehat {\widetilde \alpha}^R(t)$, the estimator  $\widehat {\widetilde \alpha}^R(t)$ is defined by a local weighted least squares criterion.
We understand the integral with respect to $\mathrm d\widetilde {N}_i$  as a Stieltjes integral.

Let $0= t_1 <...< t_{m}= T$ be an equidistant partition of the interval $[0,\mathcal T]$ with bin-width $h$ and some integer $m$. 
For $t\in[t_l,t_{l+1})$, we set
\begin{equation*}
 W^H_h(s,t)=I\{s\in[t_l,t_{l+1})\}.
\end{equation*}
The first order condition minimizing \eqref{ls:hazard} under the weighting $ W^H_h(s,t)$
leads to the histogram estimator 
\begin{equation*}
\widehat{\widetilde  \alpha}^R_{H,h}(l)= \frac{\sum_{i=1}^n\int_0^{\mathcal T} I\{s\in[t_l,t_{l+1})\}
 \, \mathrm {d}\widetilde N^R_i(s)}{\sum_{i=1}^n\int_0^{\mathcal T}  I\{s\in[t_l,t_{l+1})\}\widetilde Y_i^R(s)\, \mathrm {d}s}, \quad l=1,\dots, m.
\end{equation*}
Analogue to \cite{Hiabu:17} it can be shown that, up to lower order terms, chain-ladder development factors equal $1+\widehat{\widetilde  \alpha}^R_{H,h}(l)$.
Therefore, the following proposition can also be interpreted as a central limit theorem for development factors.
We make the following assumptions.
\begin{itemize}
\item[{[S1]}]\textit{
The bandwidth $h=h(n)$ satisfies $h\rightarrow 0$ and $n^{1/4}h\rightarrow \infty$ for $n \rightarrow \infty$.}
\item[{[S2]}]\textit{
The density $f_T$ is two times continuously differentiable}.
\item[{[S3]}]\textit{The function $l(t)=E[Z_1| \ Y_1^R(t)=1]$  is continuously differentiable.}
\end{itemize}


We define the following quantity.
\[
\gamma(t)=S_T^R(t)F_U(\min(\mathcal U, t)).
\]
\begin{proposition}\label{prop:dv}
Under Assumptions [M1]--[M3], [CLM1], [CLM2], and [S1]--[S3], for $t\in(t_l,t_{l+1})$, $ n\rightarrow \infty$, 
it holds
\begin{equation*}
(nh)^{1/2}\left\{\widehat{\widetilde  \alpha}^R_{H,h}(l)- \widetilde \alpha^R (t)-  B_H(t)\right\} \rightarrow N\left\{0,\sigma_H^2(t)\right\}, 
\end{equation*}
in distribution,
where
\begin{align*}
B_H(t) &=\left\{(t-(t_l+ \frac {1} {2} h)\right\} {{\widetilde {\alpha}^R}{}^\prime}(t)+\frac {1} {2} \left\{(t-t_l)^2+h(t_l-t)+\frac 13 h^2 \right\} {\widetilde \alpha^{R}}{}^{\prime\prime }(t), \\
\sigma_H^2(t)&= \left\{\frac {E\left[Z | \ T^R=t \right]}{ E[Z]}\right\}^2\alpha^R(t){\gamma(t)}^{-1}.
\end{align*}
\end{proposition}

\begin{proof}
See Appendix \ref{proof:prop4}.
\end{proof}

Hence, apart from the usual regularity condition,  under [CLM1] and [CLM2], consistency of the  development factors  is  achieved if both the number of observations, $n$, goes to infinity and the level of aggregation, $h$, tends to zero.

In the next section, we propose two improvements to chain-ladder development factors.
Firstly, we replace the histogram weighting, $W_h(\cdot,\cdot)$, with local polynomial kernel smoothers leading to a reduced bias. Secondly, 
we will work with the density function  instead of the hazard function because we expect estimation of the density to be more robust.
This is because the hazard function, due to the bounded support, usually
increases heavily at the right boundary whereas the shape of the density is less explosive.
A simulation study in \cite{Bischofberger:etal:19a} confirms this heuristic.

\section{Local polynomial density estimation} \label{sec:local:polynomial}
In this section we  introduce two nonparametric estimators of the one-dimensional cost-weighted density $\widetilde f_T$: the local constant estimator and the local linear estimator. 
The idea of local polynomial fitting is quite old and might originate from early time series analysis \citep{Macaulay:31}. It has been adapted to the regression case in \cite{Stone:77} and \cite{Cleveland:79}. A general overview of local polynomial fitting can be found in \cite{Fan:Gijbels:96}. 

Note that $\widetilde f_U$ can be estimated analogously by inverting the roles of $T$ and $U$ and adapting the definitions of $N_i$, $Y_i$ etc. The joint cost-weighted density $\widetilde{f}_{T,U}$ is then estimated by $\widehat{\widetilde f}_{T} \times \widehat{\widetilde f}_{U}$ in line with Remark 3. \\

We first define the cost-weighted  Kaplan-Meier product-limit estimator of the survival function $\widetilde S_T^R(t)=\int_t^{\infty}\widetilde f^R_T(s)\mathrm ds =\{E\left[Z | \ T^R \geq t \right] / E[Z]\} \int_t^{\infty} f^R_T(s) \mathrm ds$ as 
\begin{equation}
\widehat{\widetilde S}{}_T^R(t) = \prod_{s\leq t}\left \{1-\Delta \widehat{\widetilde A}{}^R (s)\right\}, \label{kaplanmeier}
\end{equation}
where 
$
\widehat{\widetilde A}{}^R(t)=\sum_{i=1}^n \int_0^t Z_i\left\{\sum_{j=1}^n Z_jY_j^R(s)\right\}^{-1} \mathrm{d}N^R_i(s)
$
 is motivated by the Aalen estimator, estimating $\widetilde A^R(t)=\int_0^t E\left[Z |\  T^R=s\right] \{ E\left[Z | \ T^R \geq s \right]\}^{-1} \alpha^R(s) \mathrm ds$. Here the product can be understood as simple finite product because of the finite number of jump points of $\widehat{\widetilde A}{}^R(t)$ as explained in \cite[p.~89]{Andersen:etal:93}. 
Let $q_p(z)=\sum_{i=0}^p \theta_iz^i$ denote a polynomial of degree $p$.
For $t\in [0,\mathcal T]$, we define the local polynomial estimator of degree $p$,  $\widehat {\widetilde f}_T^{R,p,h}(t)$ of $\widetilde f^R_T(t)$  as the minimizer  $\widehat \theta_0$ in the equation
\begin{align}\notag
\begin{pmatrix} \widehat \theta_0 \\ \vdots \\ \widehat \theta_p \end{pmatrix}
=\arg\min_{ \theta \in \mathbb R^{p+1}} \lim_{\varepsilon \downarrow 0}\sum_{i=1}^n \int &\left[\left\{ \frac 1 \varepsilon \int_{s}^{s +\varepsilon} 
 \widehat {\widetilde S}{}_T^R(w)  \, \mathrm{d}\widetilde N^R_i(w) -   q_p(t-s)\widetilde Y^R_i(s)\right\}^2 -\xi_f(\varepsilon)\right]\\  &\times K_h(t-s) \frac 1 {\widetilde Y^R_i(s)} \, \mathrm {d}s. \label{loclineq}
\end{align}
For  a kernel $K$ and bandwidth $h >0$, we set $K_h(t)=h^{-1}K(t/h)$ as usual. 
The expression $\xi_f(\varepsilon)=\{\varepsilon^{-1} \int_{s}^{s+\varepsilon}   \widehat {\widetilde S}{}_T^R(w) \mathrm{d} \widetilde N^R_i(w)\}^{-2}$ is needed to make \eqref{loclineq} well defined. 
Since $\xi_f(\varepsilon)$ does not depend on $q_p$,  $\widehat \theta_0$ is defined by a local weighted least squares criterion.

In the sequel we will only consider the cases $p=0,1$, i.e., the local constant and local linear case.
While a higher degree in conjunction with higher order kernels improves the asymptotic properties, finite sample studies show that improvements are only visible with unrealistically big sample sizes.
In the local constant case of   \eqref{loclineq} we derive the first order condition
\[
2 \theta  \sum_{i=1}^n \int_0^{\mathcal T} K_h(t-s)\widetilde Y^R_i(s)\mathrm ds = 2\sum_{i=1}^n \int_0^{\mathcal T} K_h(t-s) 
\widehat {\widetilde S}{}_T^R(s)  
\mathrm d \widetilde N^R_i(s),
\]
and conclude the local constant estimator
\begin{equation*}
\widehat{ \widetilde f}_T^{R,0,h}(t)= \frac{\sum_{i=1}^n\int_0^{\mathcal T}  K_{h}(t-s)\widehat {\widetilde S}{}_T^R(s)
\, \mathrm {d}\widetilde N^R_i(s)}{\sum_{i=1}^n\int_0^{\mathcal T} K_{h}(t-s) \widetilde Y^R_i(s)\, \mathrm {d}s}.
\end{equation*}
The final estimator in non-reversed time is then simply defined as \[
\widehat {\widetilde f}_T^{0,h}(t) = \widehat {\widetilde f}_T^{R,0,h}(\mathcal T - t) . \]
We add the following assumption.
\begin{itemize}
\item[{[S4]}] \textit{The kernel $K$ is symmetric, has bounded support and has finite second moment, and it holds $\int K(u) \mathrm du = 1$. }
\end{itemize}
Other kernels can also be used but they will require a more complex estimator.
Moreover, we introduce the  following notation.
For every kernel $K$ and $j\geq 0$,  let
\[
\mu_j(K)=\int_0^{\mathcal T} s^j K(s) \mathrm ds, \quad  
R(K)=\int_0^{\mathcal T} K^2(s) \mathrm ds. 
\]

\begin{proposition}\label{prop:2}
Under Assumptions [M1]--[M3], [CLM1], [CLM2], and [S1]--[S4], for $t\in(0,\mathcal T)$, $ n\rightarrow \infty$, it holds
\begin{equation*}
(nh)^{1/2}\left\{\widehat {\widetilde f}_T^{0,h}(t)- \widetilde f_T(t)-  B_0(t)\right\} \rightarrow N\left\{0,\sigma_0^2(t)\right\}, 
\end{equation*}
in distribution, where
\begin{align*}
B_0(t) &=h^2\mu_2(K) \left[\frac 1 2 \widetilde f^{\prime\prime}_T(t) + \widetilde f^{\prime}_T(t)  \frac{\{l(\mathcal T - t)\gamma(\mathcal T - t)\}'}{l(\mathcal T - t)\gamma(\mathcal T - t)}\right], \\
\sigma_0^2(t)&= \left\{\frac {E\left[Z | \ T=t \right]}{ E[Z]}\right\}^2R( K) f_T(t) S_T(t) {\gamma(\mathcal T -t)}^{-1}.
\end{align*}
\end{proposition}
\begin{proof}
See Appendix \ref{proof:prop2}.
\end{proof}

For the local linear case, we introduce the following quantities. For $t\in[0,\mathcal T]$, set 
\begin{align*}
G_j(t)&=\sum_{i=1}^n\int_0^{\mathcal T}  K_h(t-s)(t-s)^j \widehat {\widetilde S}{}_T^R(s) \mathrm d\widetilde N^R_i(s) \quad  (j=0,1),   \\
a_{j}(t)&=\sum_{i=1}^n\int_0^{\mathcal T}  K_h(t-s)(t-s)^j \widetilde  Y^R_i(s)\mathrm ds \quad  (j=0,1,2).
\end{align*}
The first order condition for $p=1$ then reads
\begin{align*}
G_0(t)=\widehat \theta_0a_0+\widehat \theta_1a_1, \\
G_1(t)=\widehat \theta_0a_1+\widehat \theta_1a_2.
\end{align*}
Hence, the solution $\widehat \theta_0$ is given by
\begin{equation}
\widehat{ \widetilde f}_T^{R,1,h}(t)= n^{-1}\sum_{i=1}^n\int_0^{\mathcal T} \overline K_{t,h}(t-s)\widehat {\widetilde S}{}_T^R(s)  \, \mathrm {d}\widetilde N^R_i(s), \label{loclin}
\end{equation}
where
\begin{equation*}
\overline{K}_{t,h}(t-s) = n\frac{a_2(t)-a_1(t)(t-s)}{a_0(t)a_2(t)-\{a_1(t)\}^2} K_h(t-s).
\end{equation*}
If $K$ is a second-order kernel, then
\begin{align*}
n^{-1}\sum_{i=1}^n\int \overline{K}_{t,h}(t-s)\widetilde Y_i^R(s)\mathrm ds&=1, \\
 n^{-1}\sum_{i=1}^n\int \overline{K}_{t,h}(t-s)(t-s)\widetilde Y_i^R(s)\mathrm ds&=0,  \\
n^{-1}\sum_{i=1}^n\int \overline{K}_{t,h}(t-s)(t-s)^2\widetilde Y_i^R(s)\mathrm d s&>0, 
\end{align*}
so that $\overline{K}_{t,h}$ can be interpreted as a second-order kernel with respect to the measure $\mu$, which is defined via $\mathrm d\mu(s)=n^{-1}\sum_{i=1}^n\widetilde Y_i^R(s) \mathrm ds$. 

The local linear estimator in non-reversed time is defined as 
\[
\widehat {\widetilde f}_T^{1,h}(t) = \widehat {\widetilde f}_T^{R,1,h}(\mathcal T - t). 
\]

\begin{proposition}\label{prop:3}
Under Assumptions [M1]--[M3], [CLM1], [CLM2], and  [S1]--[S4], for $t\in(0,\mathcal T)$, $ n\rightarrow \infty$, it holds  
\begin{equation*}
(nh)^{1/2}\left\{\widehat{\widetilde f}_T^{1,h}(t)- \widetilde f_T(t)-  B_1(t)\right\} \rightarrow N\left\{0,\sigma_1^2(t)\right\}, 
\end{equation*}
in distribution,
where
\begin{align*}
B_1(t) &=\frac 1 2 h^2\mu_2(K)\widetilde f_T^{\prime \prime}(t), \\
\sigma_1^2(t)&= \left\{\frac {E\left[Z | \ T=t \right]}{ E[Z]}\right\}^2R( K) f_T(t) S_T(t) {\gamma(\mathcal T -t)}^{-1},
\end{align*}
for $R(K)=\int K^2(s) \mathrm ds$.
\end{proposition}
\begin{proof}
See Appendix \ref{proof:prop3}.
\end{proof}

One alternative to estimate the cost-weighted density is to use a semiparametric asymmetric kernel density estimator which better accounts for the tail \citep{Gustafsson:etal:09}. We chose not to do so in this paper, since a nonparametric estimation technique is more in the spirit of the chain-ladder technique as explained in the previous section.

\section{Data Application: Estimating outstanding liabilities} \label{sec:application}
We apply our estimator on a data set from a motor insurance in Cyprus which was collected between 2004 and 2013. 
The data contains $n = \,$ 51,216 closed claims $(T_i, U_i, Z_i)$, $i=1,\dots,n$ 
consisting of their payment delay until the final payment $T_i$, their accident dates $U_i$, and the total claim amount $Z_i \geq 0$. 
First, we estimate the marginal cost-weighted densities $\widetilde f^T$ and $\widetilde f^U$ of $T$ and $U$, respectively, and in particular we forecast the outstanding claim amount $r_n$ consisting of all claims for accidents that have already incurred but have not been paid yet (see Remark 3 in Section \ref{sec:math:framework}). 
Afterwards, we illustrate our model assumptions [CLM1] and [CLM2] on the data set.

\subsection{Estimation and forecasting}
For the estimation of outstanding liabilities, we calculate the components  $\widehat {\widetilde f}^{0,h}=\widehat {\widetilde f}_T^{0,h}\widehat{\widetilde f}_U^{0,h}$ and  $\widehat {\widetilde f}^{1,h}=\widehat{\widetilde f}_T^{1,h}\widehat{\widetilde f}_U^{1,h}$ using the Epanechnikov kernel $K(s)=0.75 (1-s^2)I(|s|\leq 1)$. 
For data-driven bandwidth selection, we use cross-validation 
\citep{Rudemo:82, Hall:83, Bowman:84}. The score function $Q_T^j$ is motivated by the minimization problem which lead to the local polynomial estimators introduced in Chapter \ref{sec:local:polynomial}. For the estimation of $\widetilde f_T^R$ we want to minimize $\sum_{i=1}^n\int_0^1 \{ \widehat{ \widetilde f}_T^{R,j,h}(t)  - \widetilde f_T^R(t) \}^2 \widetilde Y^R_i(t) \mathrm dt $ in $h$ for $j=0,1$. Since $\widetilde f_T$ is unknown, we select the bandwidth $h_{j}^T$ as the minimizer of 
\[
\hat Q_{j}^T(h) =  \sum_{i=1}^n \int_0^{1} \left( \widehat{ \widetilde f}_T^{R,j,h}(t)\right)^2 \widetilde Y^R_i(t) \mathrm dt -  2  \sum_{i=1}^n \int_0^1  \widehat{ \widetilde f}_{T,[i]}^{R,j,h}(t) \widehat{\widetilde S}{}^R(t) \mathrm d\widetilde N^i(t)
\]
in $h$ instead \citep{Nielsen:etal:09}. The ``leave-one-out'' terms are given as 
\begin{align*} 
\widehat{ \widetilde f}_{T,[i]}^{R,0,h}(t) &= \left\{\sum_{k \neq i} \int_0^1 K_{h}(t-s) \widetilde Y^R_k(s)\, \mathrm {d}s\right\}^{-1}\sum_{k\neq i}\int_0^1  K_{h}(t-s)\widehat{\widetilde S}{}^R(s) \, \mathrm {d} \widetilde N_k(s), \\
\widehat{ \widetilde f}_{T,[i]}^{R,1,h}(t) &= n^{-1}\sum_{k\neq i}\int_0^1 \overline K_{t,h}(t-s)\widehat{\widetilde S}{}^R(s) \, \mathrm {d} \widetilde N_k(s).
\end{align*}
While this bandwidth selection works well for the estimators of the weighted density of the accident date $U$, we get unrealistic estimates for  $\widetilde f_T$. We decided to adjust the bandwidth manually and calculated $\widehat{ \widetilde f}_T^{j,h}$  for a small bandwidth $h_T^1 =  10$ days   
for delays shorter than $1.5$ years (= 548 days)
and  we used  a large bandwidth $h_T^1=500$  days 
to estimate $\widetilde f_T$ for $t>548$ days. 
The optimal bandwidths for $\widetilde f_U$ by cross-validation are $h_U^1= 471$ and $h_U^0= 280$ days, respectively. 
We remark that a full investigation of local bandwidth selection is beyond the scope of this paper.

\begin{figure}[h!]
\centering
\subfigure[][Estimated cost-weighted density of the accident date $U$ through the local constant and local linear estimators $\widehat {\widetilde f}^{0,280}_U$ and $\widehat {\widetilde f}^{1,471}_U$, respectively, with optimal bandwidths.]{
\includegraphics[width=0.46\textwidth]{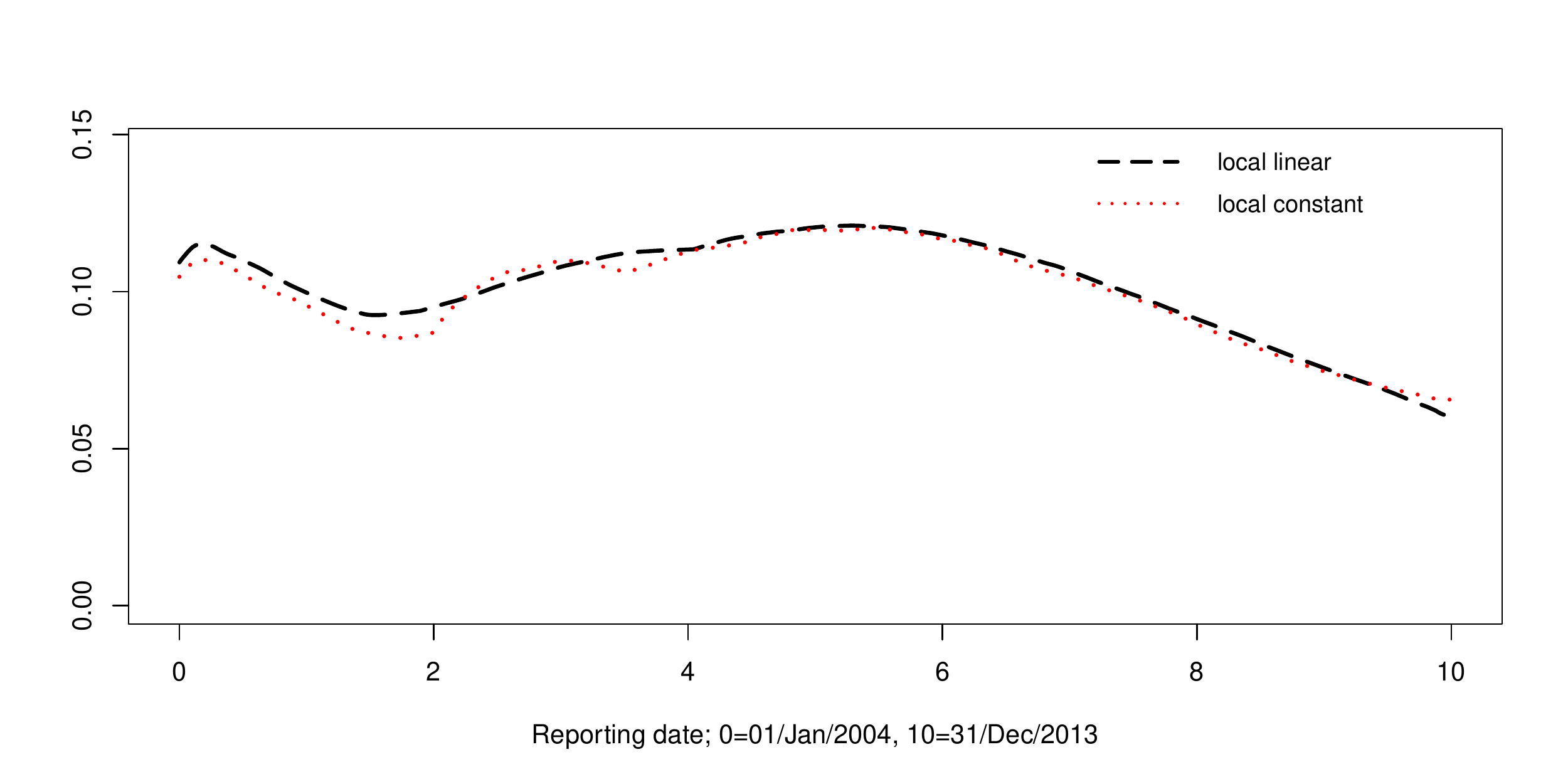}
} \ \ 
\subfigure[][Estimated cost-weighted density of the payment delay $T$ through the local constant and local linear estimators with manually corrected local bandwidths.]{
\includegraphics[width=0.46\textwidth]{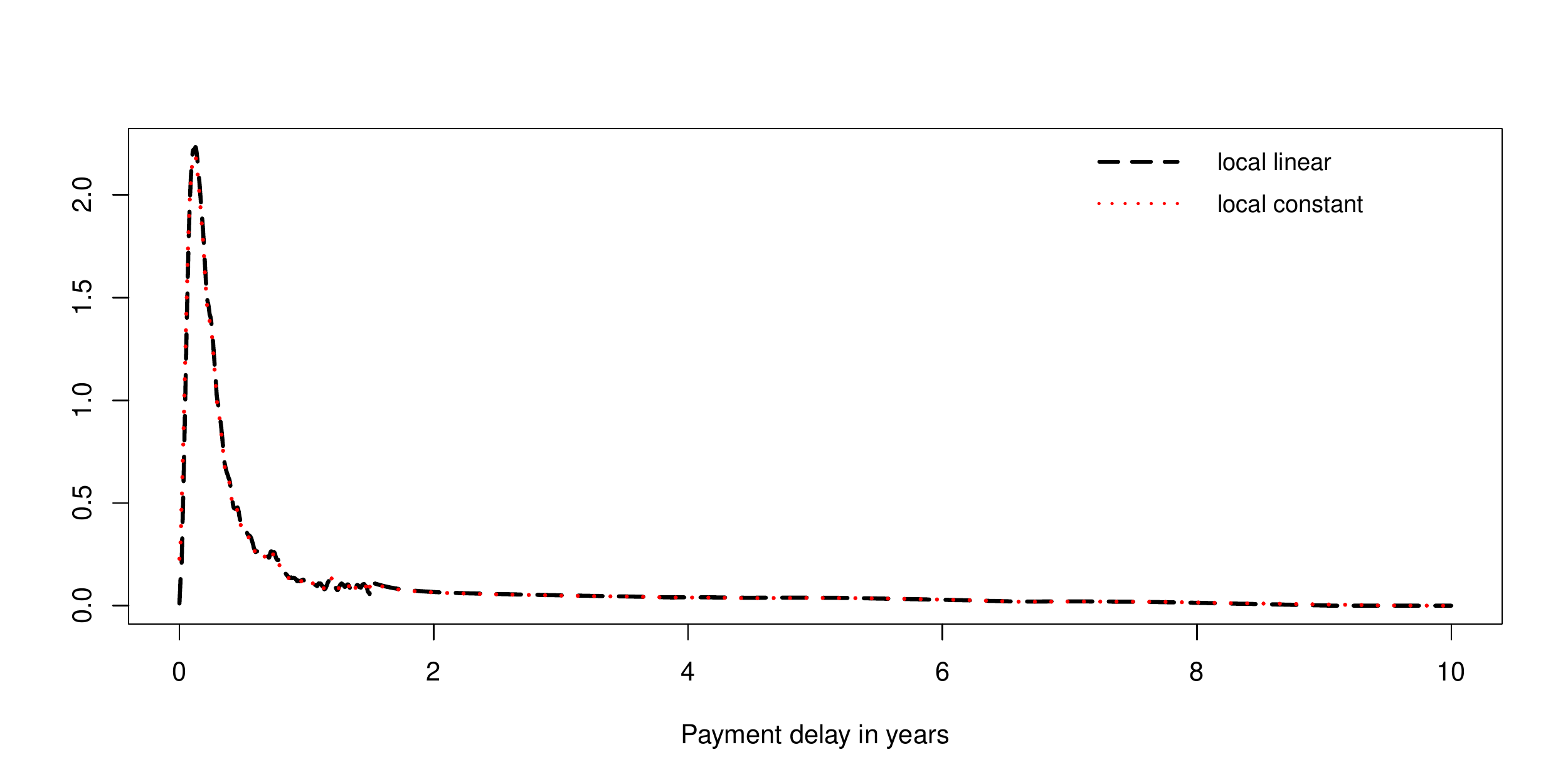}
}
\caption{Estimated marginal cost-weighted probability density functions.}
\label{fig:estimators}
\end{figure}

The results are given in Figure \ref{fig:estimators}. Since most claims were paid off after 1.5 years, our density estimators for $\widetilde f_T$ are almost zero for $t > 1.5$ years. Big outliers in that area are oversmoothed, 
which reflects the possibility of large payments with high delays better than a small number of sharp local maxima of the density at the positions of the outliers and a density of 0 elsewhere. 

For cost-weighted density estimators $\widehat {\widetilde f}_{T}$ and $\widehat {\widetilde f}_{U}$, we estimate the reserve by 
\[
\hat r_n(\widehat{\widetilde f}_{T},\widehat {\widetilde f}_{U})  =  \frac{  \int_0^{3653} \int_{3653-u}^{3653}   \widehat{\widetilde f}_{T}(t) \widehat{\widetilde f}_{U}(u)\mathrm dt \mathrm du }{\int_0^{3653} \int_{0}^{3653-u}  \widehat{\widetilde f}_{T}(t) \widehat{\widetilde f}_{U}(u) \mathrm dt \mathrm du }\sum_{i=1}^n Z_i. 
 \]
 The reserve estimate $\hat r_n$ is motivated by the representation of the reserve $r_n$ in equation \eqref{rn} in Remark 3. 
Estimates for outstanding claim payments per future year, per accident year, and in total are given in Table \ref{tab:reserves}. 
We compare the estimators with local bandwidth correction with the results obtained through the classical chain-ladder method with quarterly aggregated data.
Whereas all three total reserve forecasts are very similar, one can see differences for very short and very large delays. 
Furthermore, it is striking that both smoothed estimators forecast a non-zero claim amount for 2023 but chain-ladder estimates it to be 0. The difference between chain-ladder and smoothed density estimators for very short and mainly for large delays has already been observed for non-cost-weighted estimators in \cite{Hiabu:etal:16}. Moreover, as explained in \cite{Hiabu:17}, chain-ladder tends to overestimate the total reserve whereas the estimate from the local linear estimator is asymptotically unbiased. 
The local constant estimator is known to suffer from bias at boundaries, i.e., weaker performance than the local linear one for very short and very large delays \citep{Fan:Gijbels:96, Wand:Jones:94}.
The undersmoothed densities estimators with bandwidths obtained from cross-validation yield similar estimates for the reserve although the shape of the density estimates is unrealistically rough for larger delays (13,030,459 in the local linear case and 13,268,768 in the local constant case). 

\AtBeginEnvironment{tabular}{\scriptsize} 
\begin{table}[ht]
\begin{tabular}{rrrrrrrrrrr r}
  \hline
 future year & 2014 & 2015 & 2016 & 2017 & 2018 & 2019 & 2020 & 2021 & 2022 & 2023 & total \\ 
  \hline
CL & 3972072 & 2997371 & 2241199 & 1613228 & 1157272 & 699522 & 401116 & 190240 & 49779 & 0 & 13321800 \\ 
LL &  4606662 & 2676251 & 1944796 & 1396172 & 940474 & 562684 & 296735 & 120538 & 19231 & 10 & 12563553 \\ 
LC & 4706341 & 2734479 & 2014402 & 1473544 & 1020410 & 639236 & 357675 & 172538 & 49401 & 3420 & 13171446 \\  
     \hline \\ \hline
      accident year & 2004 & 2005 & 2006 & 2007 & 2008 & 2009 & 2010 & 2011 & 2012 & 2013 & total \\ 
  \hline
CL &        0 & 31977 & 216901 & 496947 & 830549 & 1503785 & 1778076 & 2221387 & 2387308 & 3854869 & 13321800 \\
LL &      18 & 28133 & 179632 & 440042 & 823458 & 1323252 & 1734902 & 2055000 & 2364939 & 3614176 & 12563553 \\ 
LC & 5144 & 62617 & 248154 & 499977 & 906861 & 1403358 & 1792929 & 2100946 & 2372382 & 3779076 & 13171446 \\ 
   \hline
   \end{tabular}
\caption{Forecasted claim amount for future years and by accident year estimated by chain-ladder (CL), manually corrected local linear (LL) and local constant estimators (LC).  }
\label{tab:reserves}
\end{table}
\AtBeginEnvironment{tabular}{\footnotesize} 
We are aware that these forecasts are just point estimates for the reserve. We investigate variation in the forecast under a controlled setting in the simulation study in the next chapter.

\subsection{Illustration of assumptions} \label{subsec:ver:ass}

\begin{figure}[h!]
\centering
\subfigure[][Independence check on real data.]{
\includegraphics[width=\textwidth]{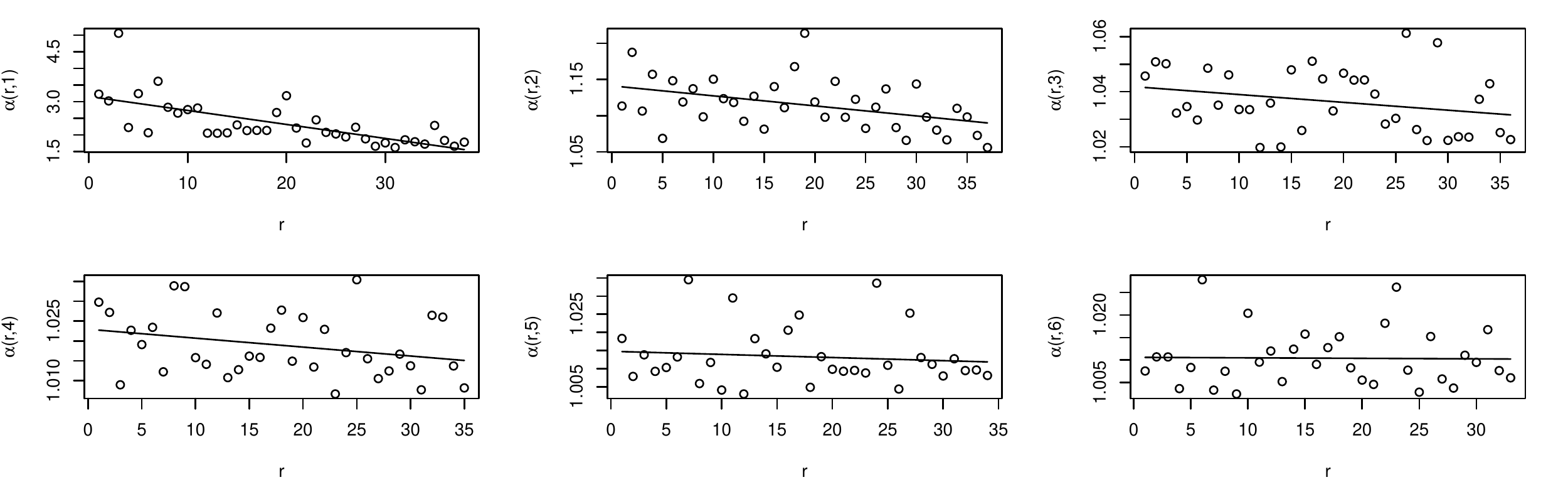} 
} \\
\subfigure[][Independence check on simulated independent data for comparison.]{
\includegraphics[width=\textwidth]{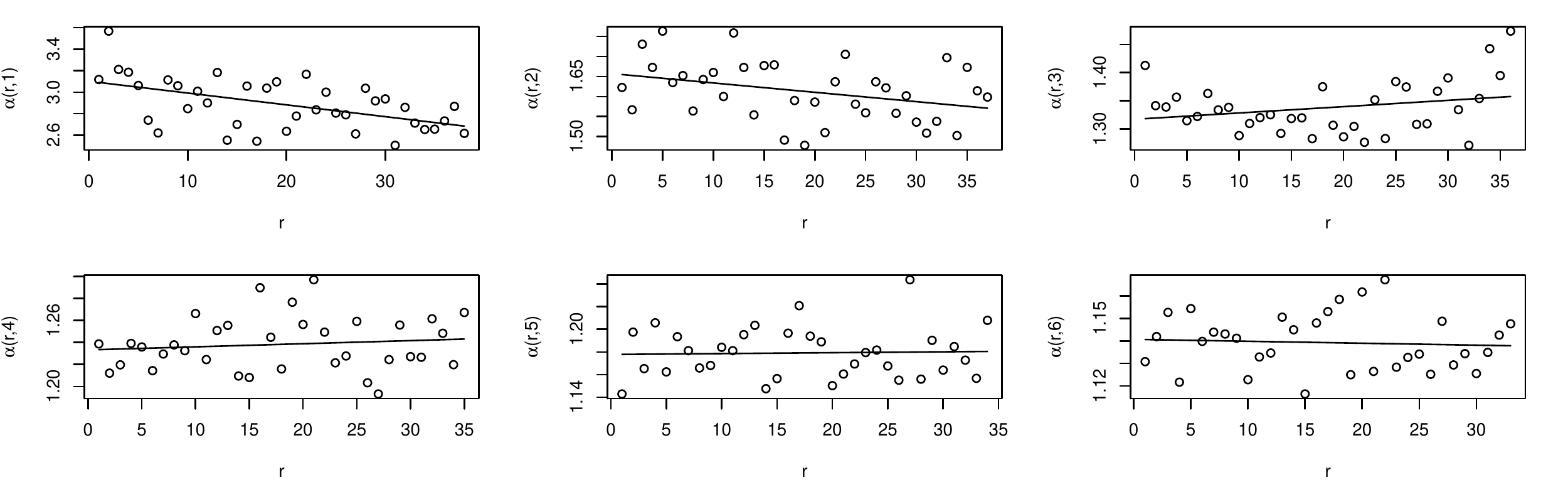}
}
\caption{Illustration of dependence in the data set (a) and in simulated independent data for comparison (b). The graphs show the development factors for the first six accident year quarters 1 = Q1 2004 to  6 = Q2 2005. For the values of $r$ it holds 1= Q1 2004, \dots, 40 = Q4 2013.}
\label{fig:check:indep}
\end{figure}

For Assumption [CLM1], the independence between $T$ and $U$ could not be assured by an independence test based on Conditional Kendall's tau for truncated data \citep{Austin:Betensky:14,Martin:Betensky:05}. 
To get more insight we aim to visualize the underlying dependency.
We aggregate the data into three-month bins $q_1,\dots,q_{40}$. Then we introduce a  triangle with aggregated observations $\mathcal{N}^Q_{r,s} = \sum_{i=1}^n I(U_i \in q_r, T_i \in q_s) $, $r,s,=1,\dots,40$, $r+s \leq 40$,  and calculate the development factors $\alpha(r,s)= \sum_{l=1}^{s+1} \mathcal{N}^Q_{r,l} / \sum_{l=1}^{s} \mathcal{N}^Q_{r,l}$, for development quarter $s$ and  accident quarter date $r$.
The values of $\alpha$ for the first six development quarters are given in Figure \ref{fig:check:indep}a. 
Under the assumption of independence between $T$ and $U$ (Assumption [CLM1]), the function $\alpha$ is independent of the accident date $r$ and hence each plot should show points scattered around a horizontal line yielding  a flat regression line.

The $p$-values for the linear regression slope parameters were only significant at 5\%-level in the first two quarters.
For comparison, Figure \ref{fig:check:indep}b shows the development factors on independently simulated variables for which linear trends were insignificant at 10\%-level in every quarter. Except for the first plot in Figure \ref{fig:check:indep}a, indeed none of the plots and linear fits from Figure \ref{fig:check:indep}a are visually distinguishable from any plot in Figure \ref{fig:check:indep}b.  
Clearly, this is not a sound method to prove independence but it illustrates that the dependence in our data might come from the first quarter only. 
Advisable extensions of our model, that handle possible dependence in our data, are  e.g.\ seasonal effects as considered in \cite{Lee:etal:15} or operational time \citep{Lee:etal:17}. We do not consider these approaches in this paper.

\begin{figure}[h!]
\centering
\subfigure[][On the real data set.]{
\includegraphics[width=\textwidth]{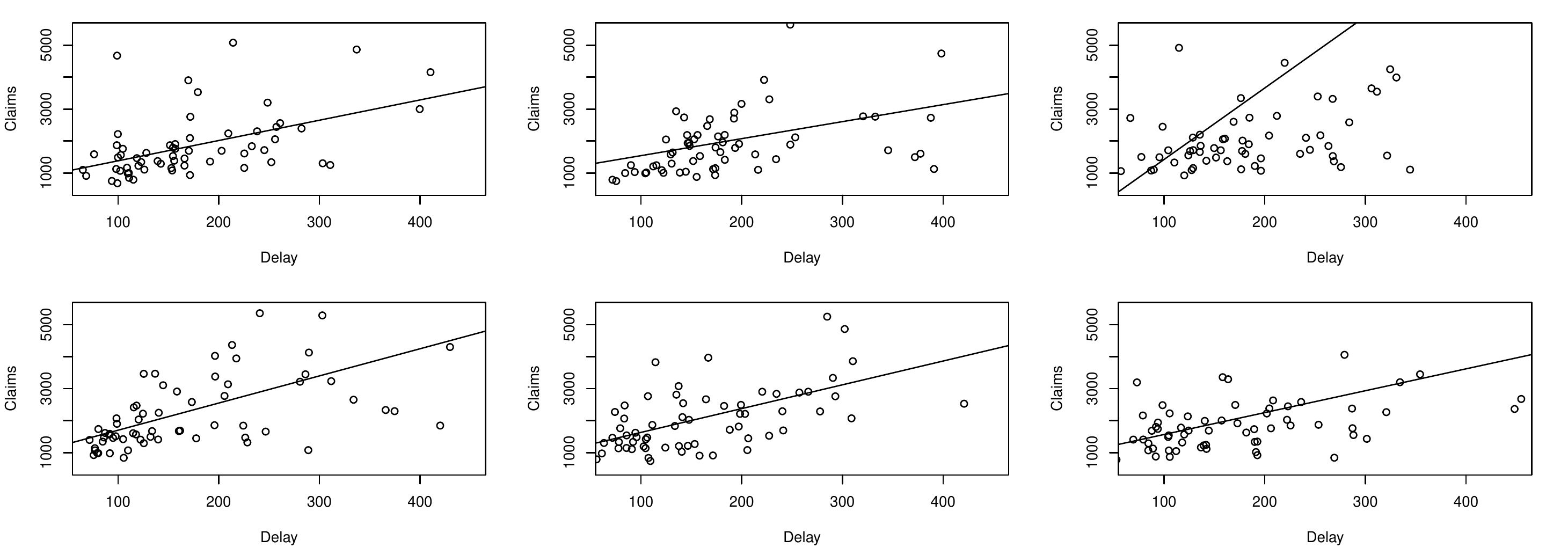}
} \\
\subfigure[][On simulated data on which Assumption {[CLM2]} is assured. ]{
\includegraphics[width=\textwidth]{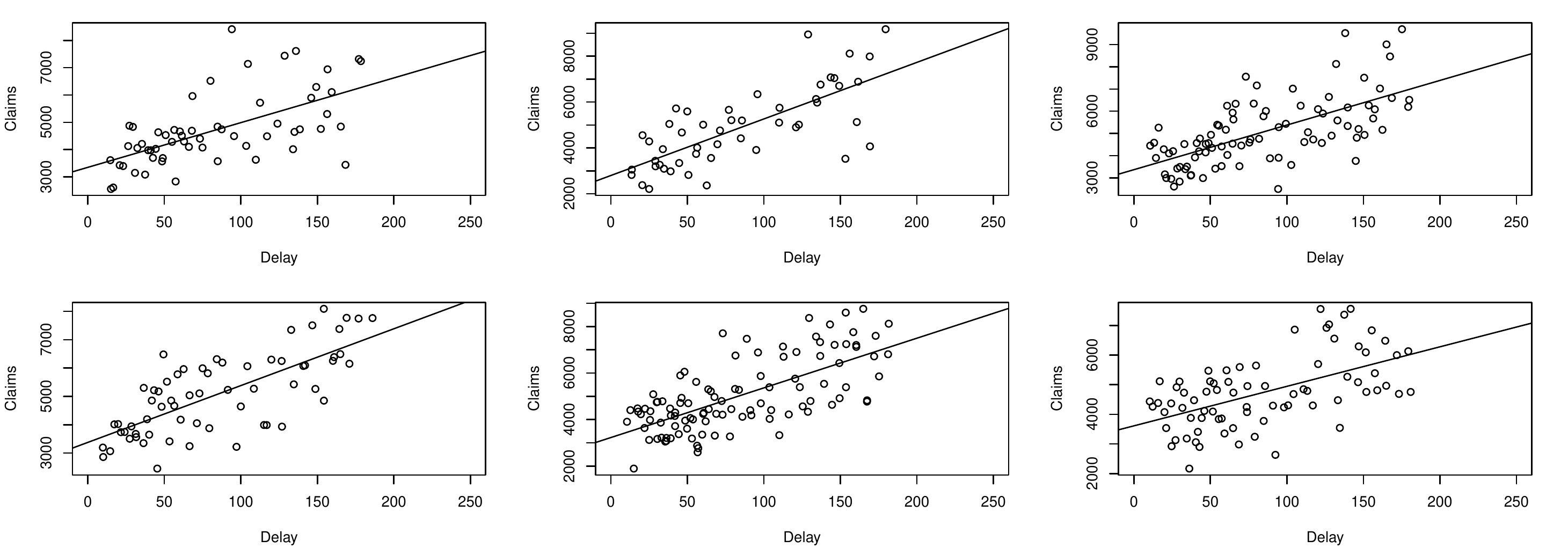}
} 
\caption{Claim costs $Z_i $ plotted against the delay $T_i$ in days for data in the first six accident year quarters Q1 2004 to  Q2 2005. }
\label{fig:check:assumption}
\end{figure}

For Assumption [CLM2], we have to verify that the expected cost conditioned on $T$ and $U$ is multiplicatively separable, i.e., that there exist functions $m_1$, $m_2$ such that $E[Z|T,U]=m_1(T)m_2(U)$. 
Similarly to the above, we use a visual approach on aggregated data to illustrate this setting in the first six quarters $q_1,\dots,q_6$ of the accident years. 
Assumption [CLM2] is satisfied if for all observations $i$ with $U_i$ in quarter $q_k$ it holds $Z_i=c_k m_1(T_i) +\varepsilon_i$ for a quarter dependent constant $c_k>0$, and a mean-zero error $\varepsilon_i$. Figure \ref{fig:check:assumption} shows the claim cost given the delay for claims in the first six quarters. 
Under Assumption [CLM2], the points in each plot 
should be generated by the same regression function after normalizing with the accident date quarter dependent factor $c_k$. 
We use a linear interpolation 
to compare the structure of the observations. 
All but the third plot show very similar development of claim costs. 
In the third plot, the claim costs increase much faster due to some outliers that are not visible in the plot. 
For comparison, we generate 50,000 observations from Scenario 5 in Section 6 where $ E[Z | T, U] = m_1(T)m_2(U)$ with $m_1(t)=(t+0.75)$, $m_2(u)= (u-0.25)^2+1$.


We conclude that while the data does not fully follow our assumptions, it is suitable enough for the illustration purpose of this paper. 

\section{Simulation study}   \label{sec:simul}
This chapter shows the performance of our new estimators on simulated data. Our first finding is that the local linear estimator outperforms the local constant one at boundaries (Section \ref{subsec:density:simulation}). Secondly, when estimating the reserve, our local constant estimator is best for small sample sizes and the local linear one for large sample sizes whereas the performance of chain-ladder varies (Section \ref{subsec:reserve:estimation}). Last, in a micro model in Section \ref{subsec:micro:simulation}, we see that reliable monthly forecasts can only be obtained with our density estimators and not with chain-ladder. 

\subsection{Weighted density estimation}\label{subsec:density:simulation}
\begin{figure}[h!]
\centering
\subfigure[][$ f^U$: Truncated mixed normal.]{
\includegraphics[width=0.45\textwidth]{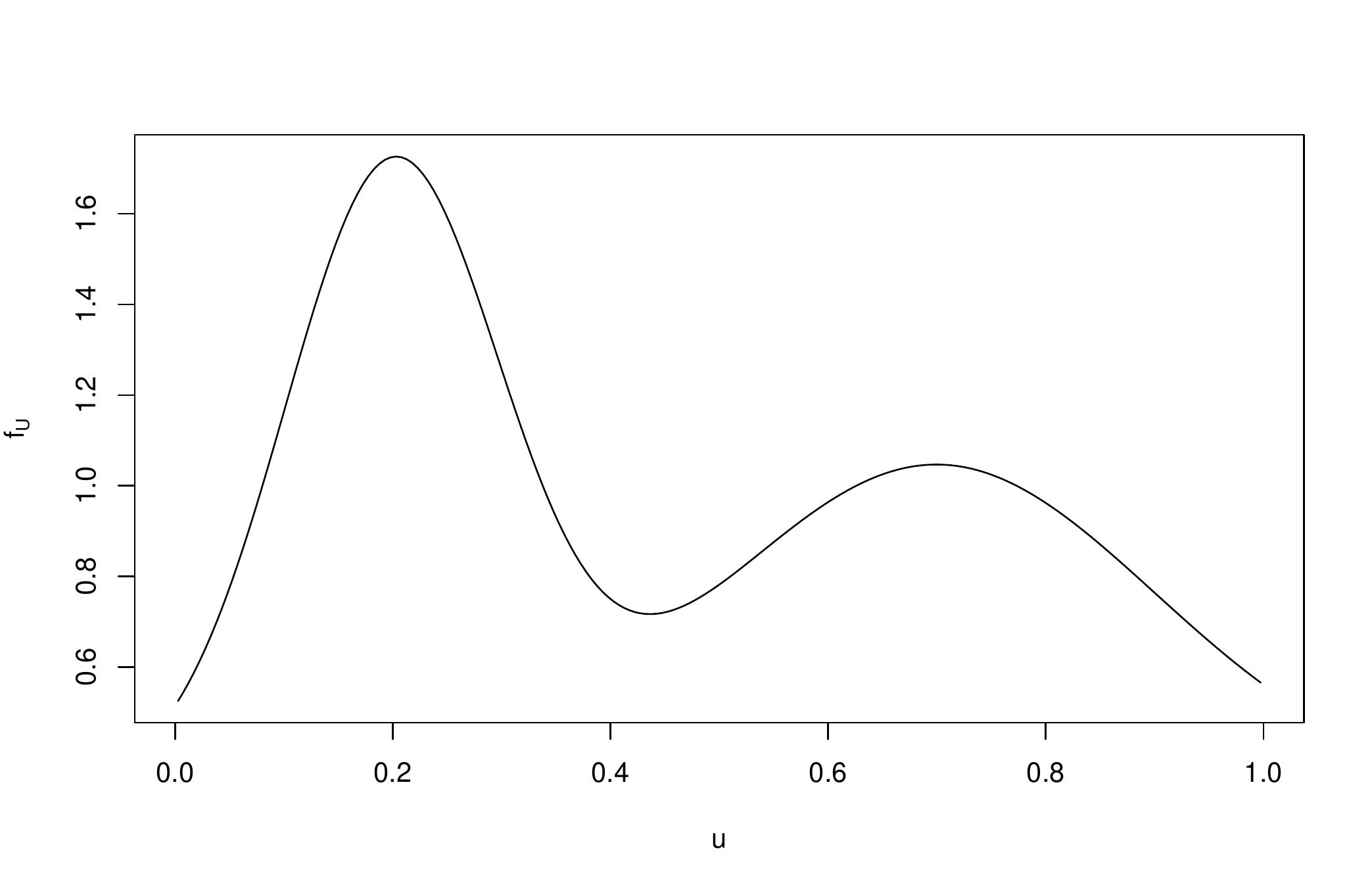}
} \ \ 
\subfigure[][$ f^U$: Boundary challenge.]{
\includegraphics[width=0.45\textwidth]{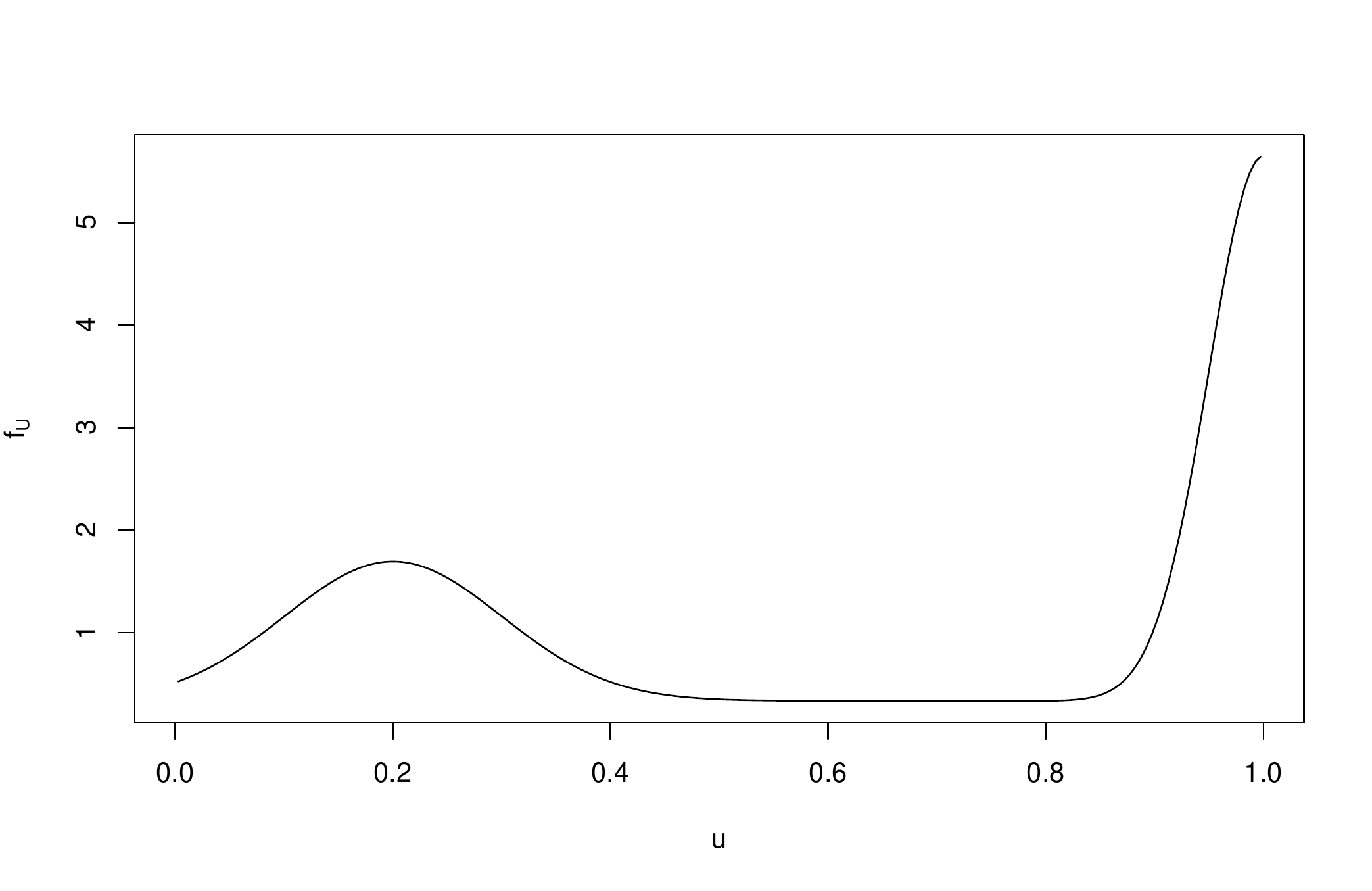}
}
\subfigure[][$ f^T$: Decreasing beta.]{
\includegraphics[width=0.45\textwidth]{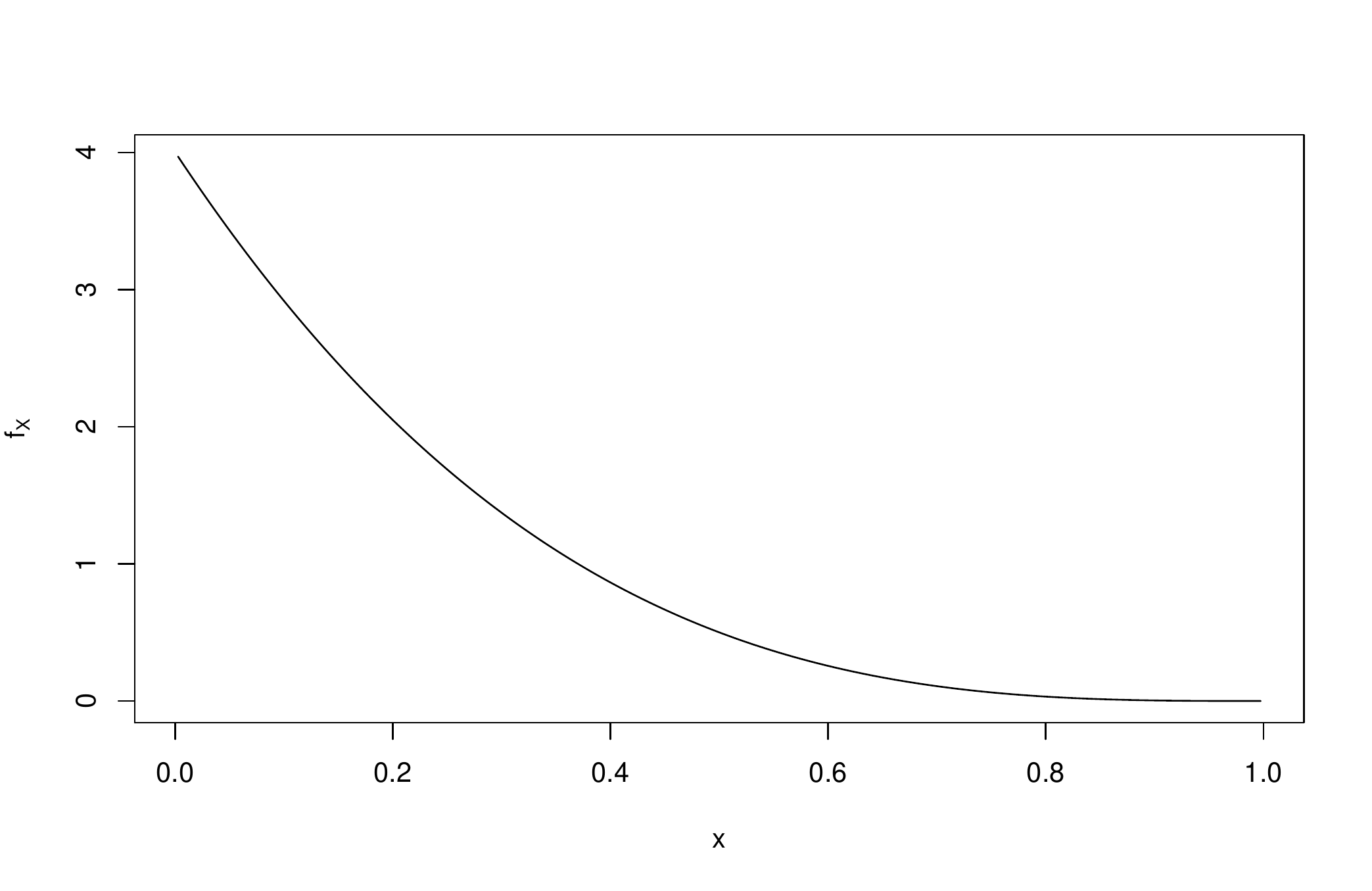}
} \ \ 
\subfigure[][$f^T$: Mixture of betas.]{
\includegraphics[width=0.45\textwidth]{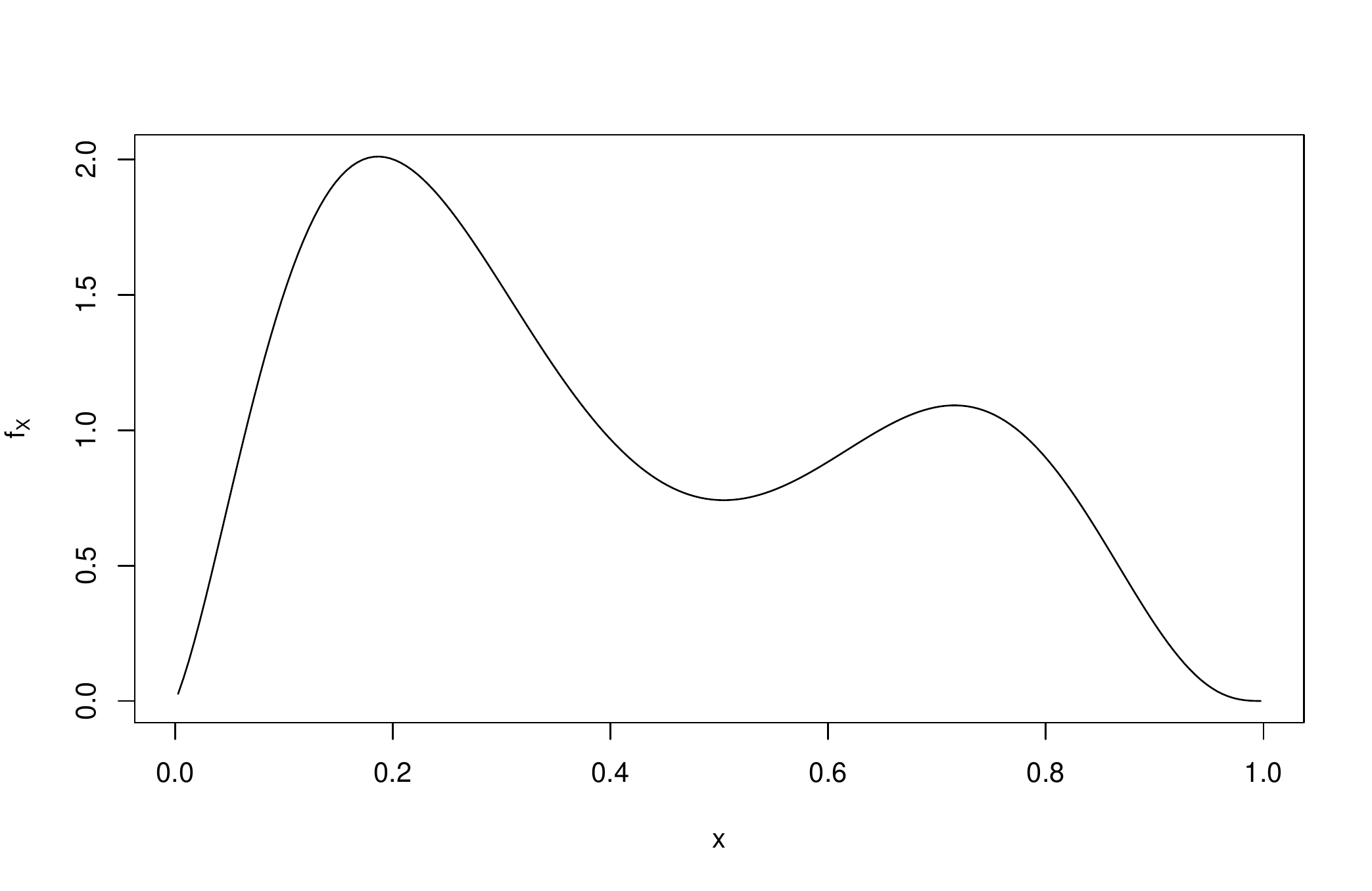}
}
\subfigure[][Moderately decreasing gamma distribution for $Z$. Histogram for 10,000 observations $(Z_i | (T_i, U_i))$, where $U_i$ is simulated from (a) and $T_i$ from (c). Bin width is 500.]{
\includegraphics[width=0.45\textwidth]{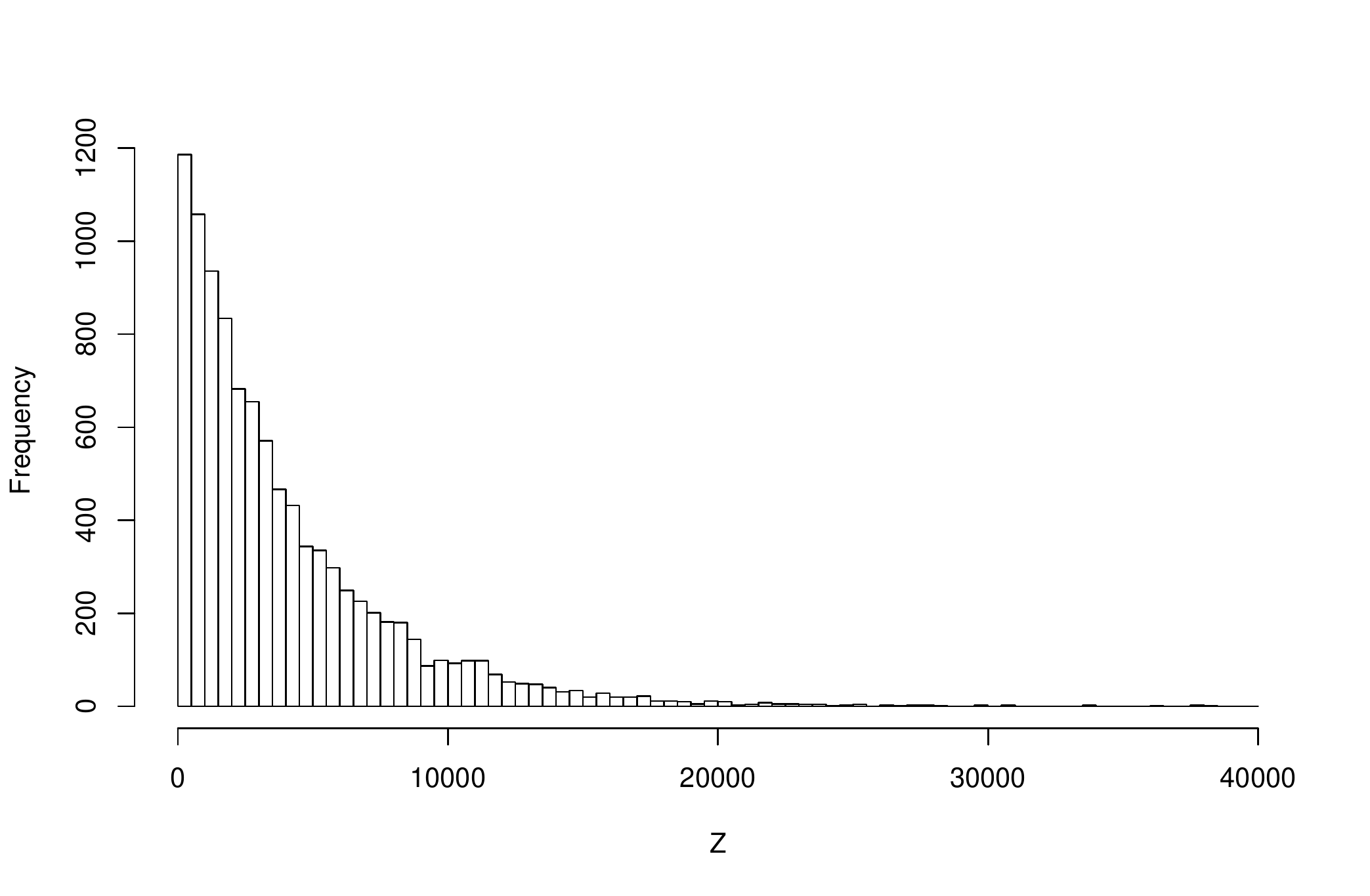}
} \ \ 
\subfigure[][Heavily decreasing gamma distribution for $Z$. Histogram for 10,000 observations $(Z_i |  (T_i, U_i))$, where $U_i$ is simulated from (a) and $T_i$ from (c). Bin width is 125.]{
\includegraphics[width=0.45\textwidth]{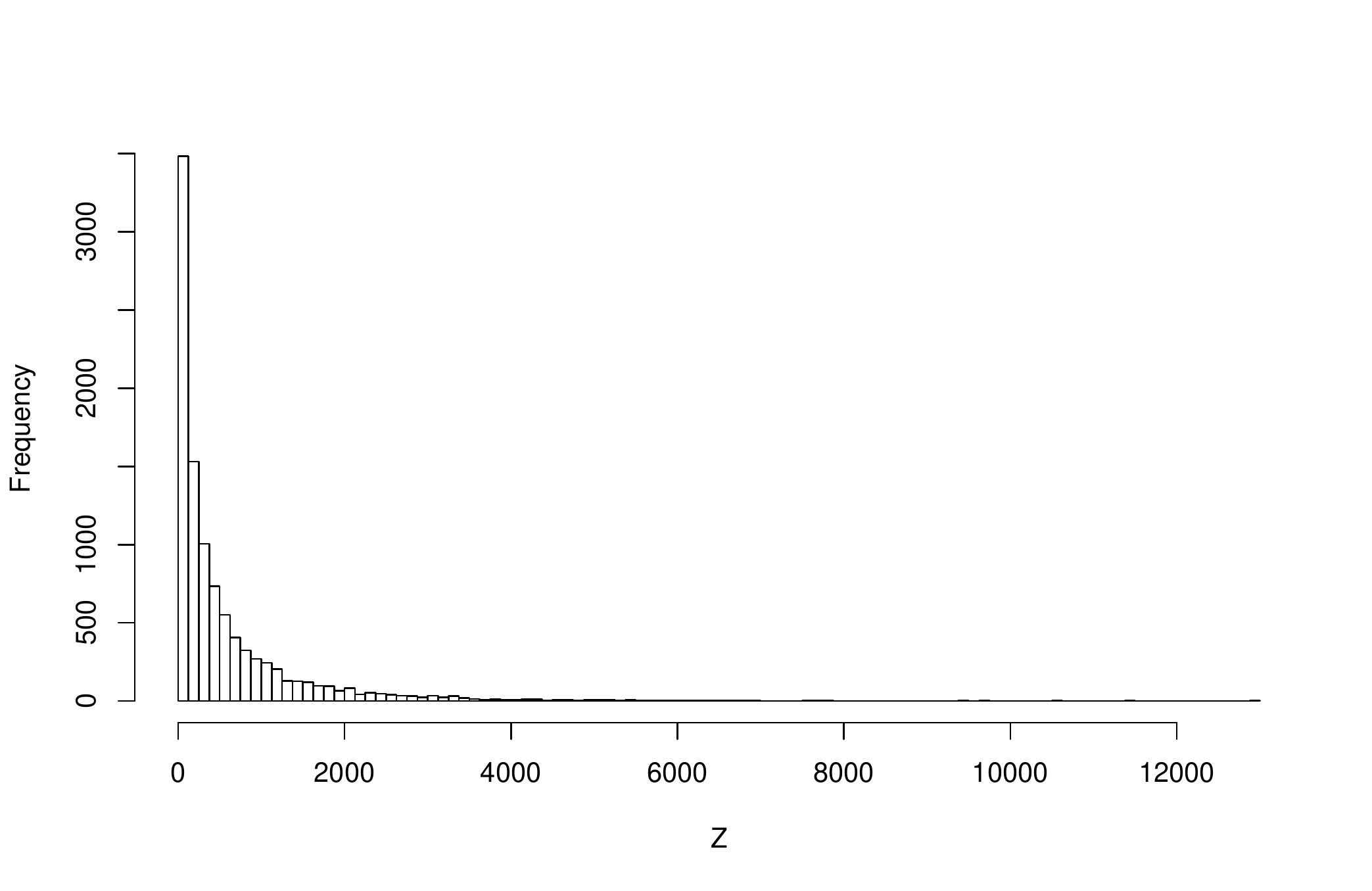}
}
\caption{Probability distribution functions and histograms of the distributions used for the simulation study.}
\label{fig:distributions}
\end{figure}

We perform a simulation study to show the performance of our estimators for a selection of distributions if the optimal bandwidth is chosen. We simulate truncated observations $(T_i,U_i,Z_i)$, $i=1,\dots,n$ on $\mathcal I=\{(t,u): 0\leq u,t \leq 200, u+t \leq 200\}$. 
With the true weighted densities $\widetilde f_T$ and $\widetilde f_U$  being known, we calculate the local constant and local linear estimators $\widehat {\widetilde f}^{j,h}_{A}$, $j=0,1$, with the best bandwidth $h^j_A$ with respect to the integrated squared error
\[
\ISE(\widehat {\widetilde f}^{j,h^j_A}_{A}, \widetilde f_A) = \int_0^1 \left( \widehat {\widetilde f}^{j,h^j_A}_{A}(t) - \widetilde f_A(t) \right)^2 \mathrm dt,
\]
for $A\in\{T,U\}$. 
We choose 
eight different settings for the distributions of $T$, $U$, and $Z$. 
The choice of the distributions is motivated by empirical distributions on the one hand and challenging estimation settings for the distribution of $T$ and $U$ are added on the other hand. The observations of $T$ and $U$ are simulated independently and truncated on $[0,1]$.    
The probability density functions for $T$ and $U$ are shown in Figure \ref{fig:distributions}a--d and the values of $Z$ given one choice of $(T,U)$ are illustrated as histograms in Figure \ref{fig:distributions}e and f. 
%
%
%
For simulated conditional claim costs $Z$ given $T$ and $U$, we take gamma distributions with shape parameter $k =1$ and different scale parameters $\theta = (T+0.75)((U-0.25)^2+1)$ and $\theta = T(U^2-U+1)$. Note that Assumption [CLM2] holds because of the identity $E[Z | T, U] = k \theta$ of the gamma distribution. 

We take all combinations of these distributions and label the eight scenarios as given in Table \ref{tab:scenarios}. For each scenario 1000 random samples of sizes 100, 1000, 10,000 and 100,000 are generated. \\

\begin{table}[h!]
\centering
\begin{tabular}{llll}
  \hline
Scenario & $T$ & $U$ & $Z$  \\ 
  \hline
1  &  decreasing beta&   truncated mixed normal & moderately decreasing \\
2 &  decreasing beta&   truncated mixed normal &  heavily decreasing \\
3  &  decreasing beta&  boundary challenge& moderately decreasing \\
4 &   decreasing beta&  boundary challenge&  heavily decreasing \\
5  &      mixture of betas& truncated mixed normal&    moderately decreasing \\
6 &    mixture of betas& truncated mixed normal&    heavily decreasing \\
7  &   mixture of betas&  boundary challenge&  moderately decreasing \\
8 & mixture of betas&  boundary challenge&    heavily decreasing \\
   \hline
\end{tabular}
\caption{Scenarios in the simulation study.}
\label{tab:scenarios}
\end{table}

We investigate 32 cases arising from eight different scenarios and four different sample sizes. 
The exact results are omitted here. We only give our main conclusion and focus on the reserve estimate in more detail in the next section. 
The local constant and the local linear estimators perform similarly in terms of empirical mean integrated squared error (eMISE) 
with the local linear estimators being more stable. 
In 26 out of 32 cases, the eMISE of $ \hat f_{1,h_1^X,K}^X$ is more than 25\% lower than the of $\hat f_{0,h_0^X,K}^X$. In 4 cases it even improves the eMISE by more than 75\%. On the other hand, there are only two cases where the local linear estimator leads to an increase in the eMISE by 0.7\% and 6.7\%, respectively. 
In the other covariate, both density estimators perform equally well. 
However, the local linear estimator performs better in scenarios with the boundary challenge distribution for $U$. This reflects aforementioned weakness of the local linear kernel density estimator close to boundary regions. The difference is biggest in Scenarios 7 and 8 where the local linear estimator is able to make up for the lack of observations in the corner. 

\subsection{Estimates for stimulated outstanding liabilities} \label{subsec:reserve:estimation}
Next, we compare the reserve estimates 
\[
\hat r_n(\widehat {\widetilde f}_{T},\widehat {\widetilde f}_{U}) =\frac{ {  \int_0^1 \int_{1-u}^{1}   \widehat {\widetilde f}_{T}(t) \widehat {\widetilde f}_{U}(u)\mathrm dt \mathrm du }}{\int_0^1 \int_{0}^{1-u}  \widehat {\widetilde f}_{T}(t) \widehat {\widetilde f}_{U}(u) \mathrm dt \mathrm du }\sum_{i=1}^n Z_i
\]
with the true outstanding claim amount $r_n$ defined in equation \eqref{rn}. 

Table \ref{tab:reserve} contains the mean, standard deviation and the median of the errors in the estimation of the squared relative errors
\[
\err^2(\widehat {\widetilde f}_T,  \widehat {\widetilde f}_U) =\left( \frac{\hat r_n(\widehat {\widetilde f}_T, \widehat {\widetilde f}_U) - r_n}{r_n} \right)^2.
\]
The results are compared to the estimation through chain-ladder applied on the triangle arising from the aggregation of the simulated observations of $T$ and $U$ into 20 bins each. This aggregation is comparable to quarterly aggregation on real data. 

\begin{figure}[h!]
\centering
\subfigure[][1,000 simulated past claims weighted by payment amount sampled from the distribution in (d).]{
\includegraphics[width=0.3\textwidth]{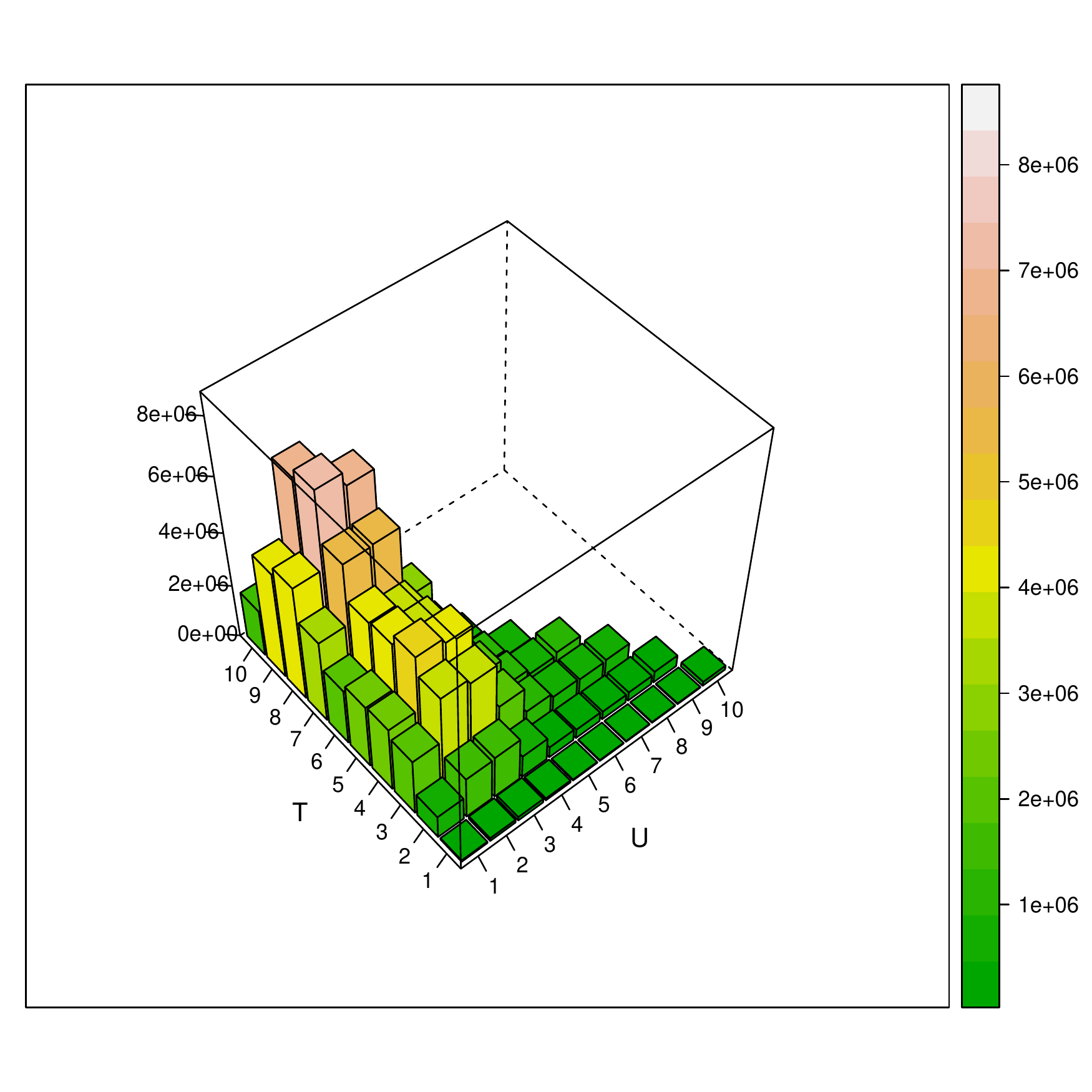}
} \ \ 
\subfigure[][Past claims and estimated outstanding claims from chain-ladder per underwriting year $U$ and reporting delay $T$.]{
\includegraphics[width=0.3\textwidth]{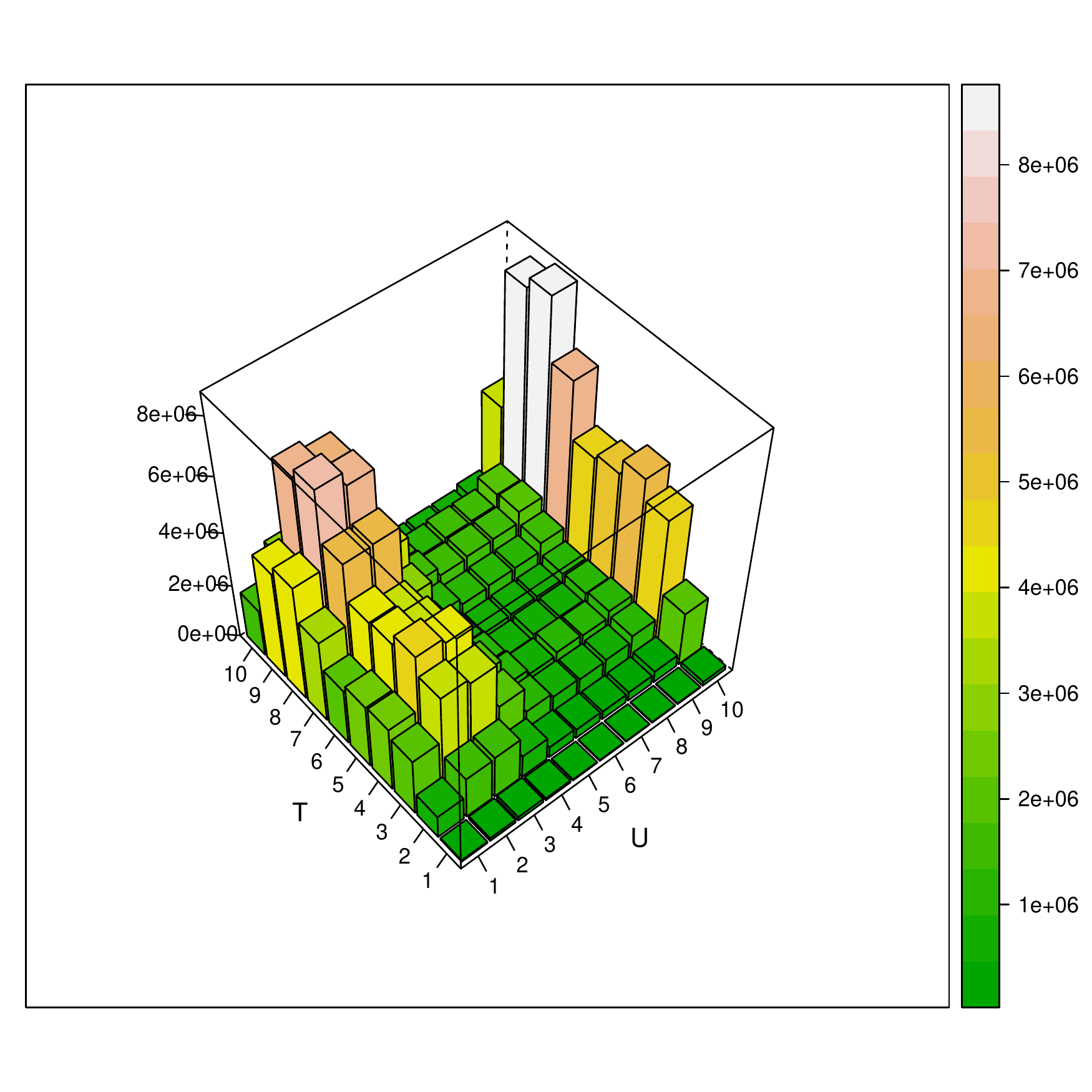}
} \ \ 
\subfigure[][Estimated outstanding claims from chain-ladder method per underwriting year $U$ and reporting delay $T$ without original data.]{
\includegraphics[width=0.3\textwidth]{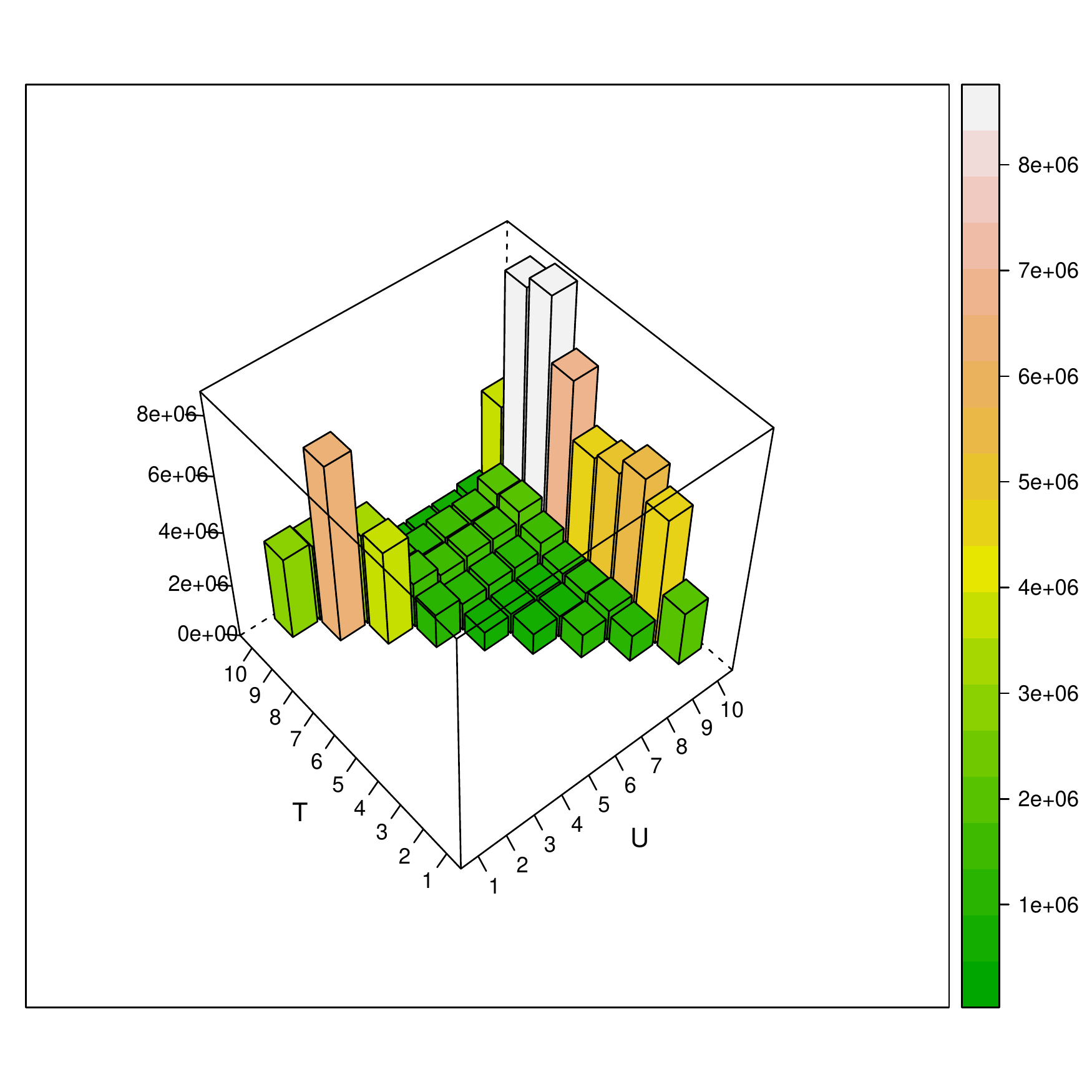}
}
\subfigure[][True density $\widetilde f_{T,U}$ of the underlying cost-weighted distribution. ]{
\includegraphics[width=0.3\textwidth]{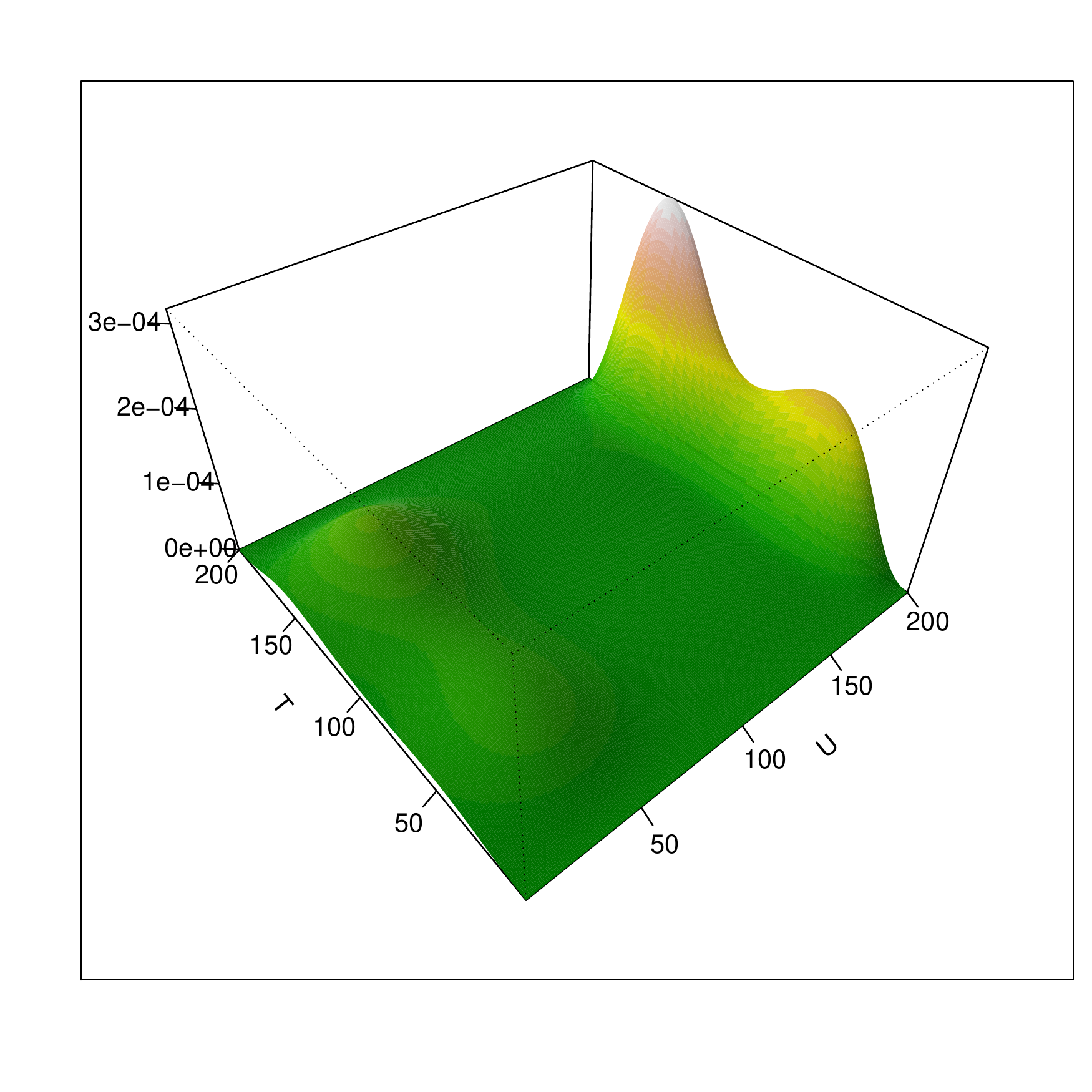}
} \ \ 
\subfigure[][Local linear estimator of the cost-weighted density $\widetilde f_{T,U}$.]{
\includegraphics[width=0.3\textwidth]{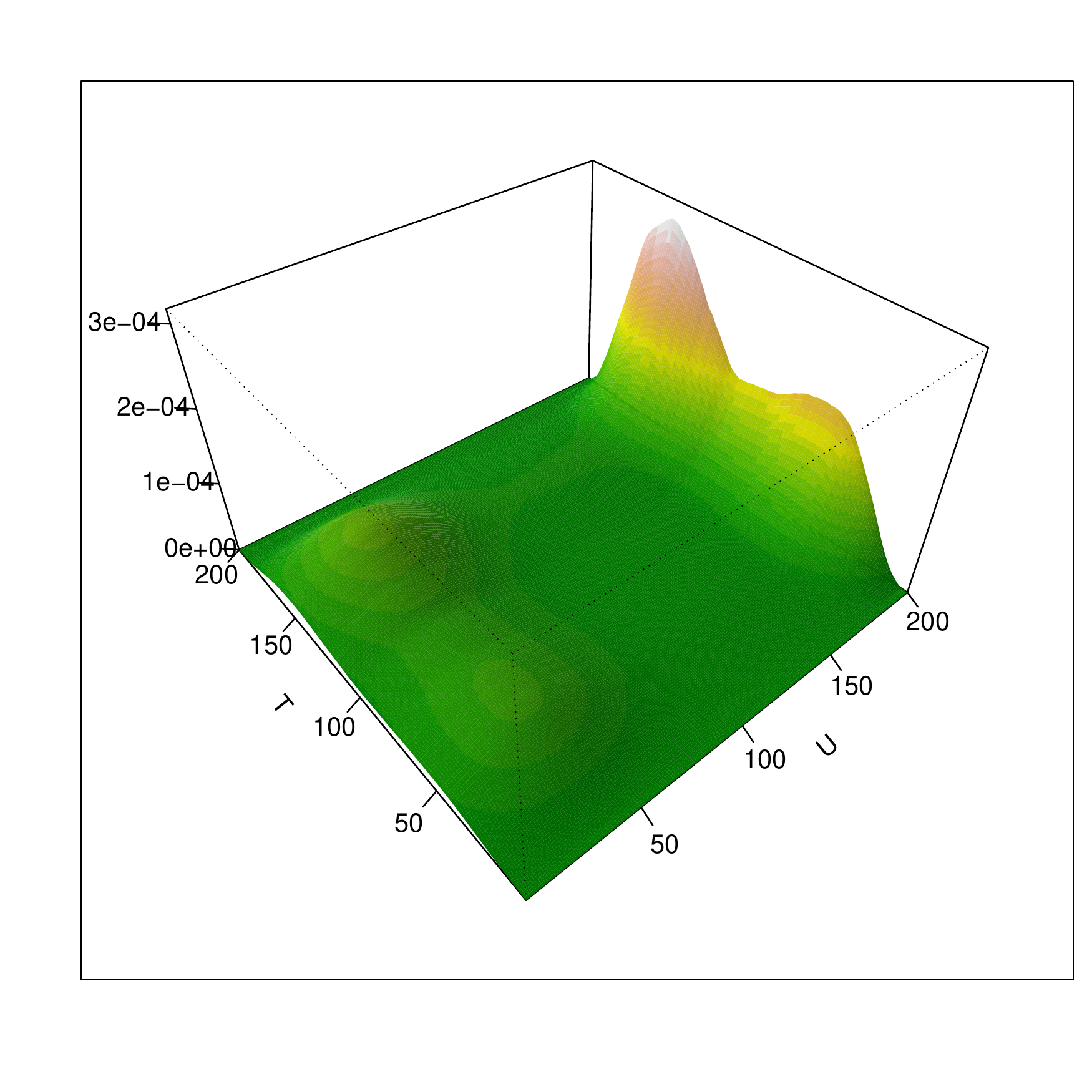}
} \ \ 
\subfigure[][Local constant estimator of the cost-weighted density $\widetilde f_{T,U}$.]{
\includegraphics[width=0.3\textwidth]{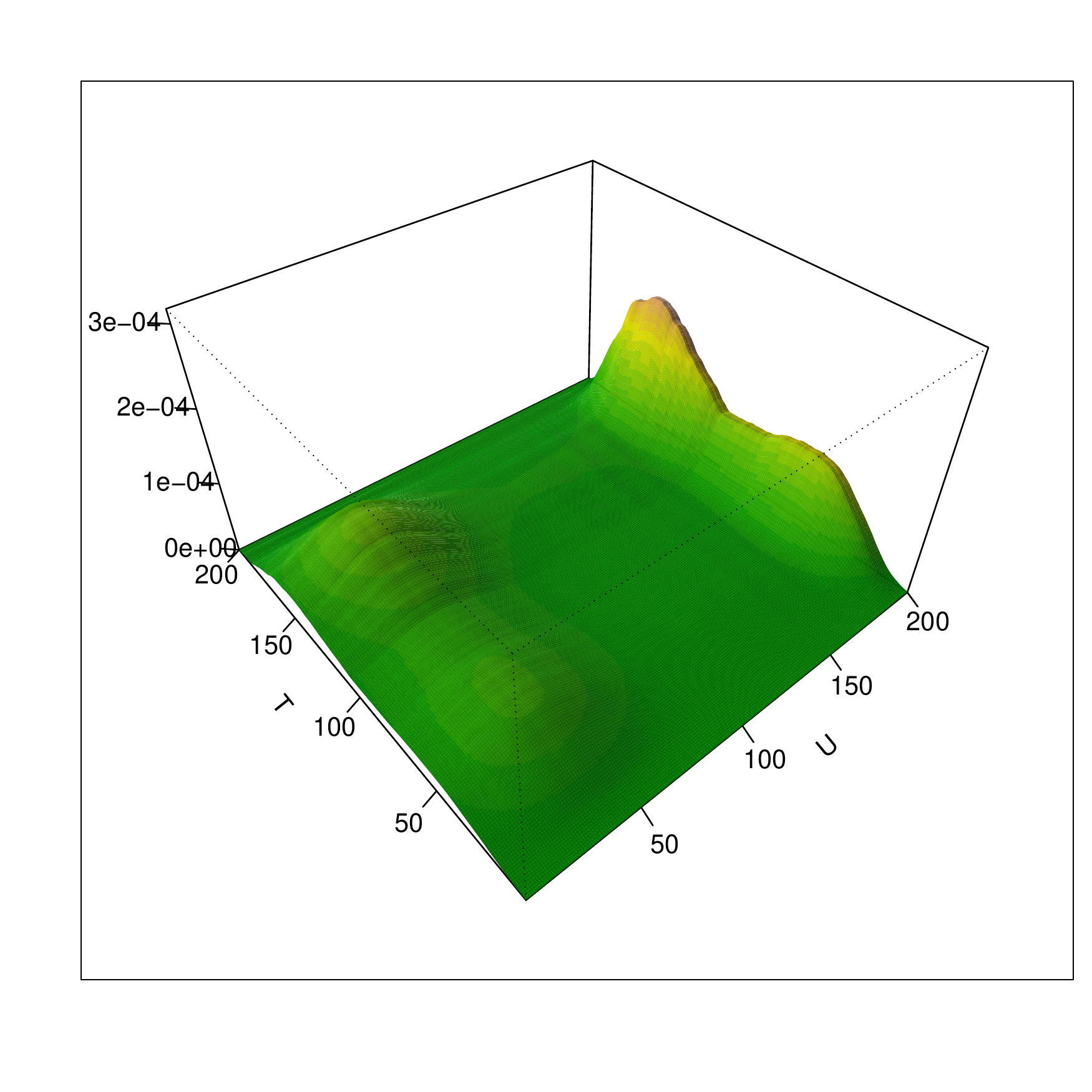}
}
\caption{Simulated claims, chain-ladder estimates, underlying distribution, local linear and local constant estimator for one simulation run of Scenario 8 with 100,000 simulated observations. 
}
\label{fig:scenario8}
\end{figure}

First, we want to note that there was a complete breakdown of the chain-ladder algorithm for too small numbers of observations which resulted in an invalid estimate in our implementation. 
%
Moreover, in most cases of the simulation study, our local polynomial density estimators outperform chain-ladder. 
%
The reserve estimates from the local linear estimators were strikingly better in the boundary challenge Scenarios 3, 4, 7  and 8 for numbers of observations larger than $n=1000$. 
For $n=100$ the local constant reserve estimate was best in six out of eight scenarios. 
%
In boundary challenge scenarios, chain-ladder not only lead to invalid results for small sample sizes but it also resulted in extreme outliers. 
An illustration of the results is given in Figure \ref{fig:scenario8}. It shows a scenario in which the local linear density estimator is the only one that estimates the altitude of the maximum in the joint density almost correctly. 

%
%

We conclude that the local linear estimator performs best for $n\geq 1000$ and that the local constant one does for smaller sample sizes. Detailed results can be found in Table \ref{tab:reserve}. 

\input{summary.res.tex}

\subsection{Simulation of a micro model} \label{subsec:micro:simulation}
In this section, we investigate the performance of our estimators and chain-ladder on simulated data arising from a micro model. 
We simulate different steps in the underwriting and payment process separately and for different types of policies in one line of business and estimate outstanding payments under different circumstances. We then compare estimates of the reserve with actual future payments. 

To create our data set, we follow the ``central scenario'' simulation in \cite{Baudry:Robert:19}. We generate mobile phone insurance policies that were underwritten over two years and estimate the outstanding liabilities at different times throughout the underwriting period and some months after the last policy was sold. We assume that the insurance provider covers damage in the three events breakage, oxidation, and theft. For this purpose the three policy types ``breakage'', ``breakage and oxidation'', and ``breakage, oxidation, and theft'' are underwritten with probabilities 0.25, 0.45, and 0.30. 
Moreover, there are four different mobile phone brands with four different models each specifying the price of the phone. 
The frequencies of brands and models and their basis prices and model prize factors are given in Table \ref{tab:brands:models}. 
\begin{table}[h!]
\centering
\begin{tabular}{lll|lll}
  \hline
Brand & Probability & Basis price & Model & Probability & Price factor  \\  
  \hline
Brand 1 & 0.45 & \$600 & 0  & 0.05 & 1 \\
Brand 2 & 0.30 & \$550 &  1 & 0.10 & 1.15 \\
Brand 3 & 0.15 & \$300 & 2 & 0.35 & $1.15^2$\\
Brand 4 & 0.10 & \$150 & 3 & 0.50 & $1.15^3$\\
   \hline
\end{tabular}
\caption{Distribution, basis price, and price factor of phone brands and models.}
\label{tab:models}
\label{tab:brands:models}
\end{table}

Following \cite{Baudry:Robert:19}, we simulate insurance policies that are underwritten independently between the first day of 2016 and the last day of 2017. 
The number of underwritten policies per day follows a Poisson point process with constant intensity $\lambda_0(t) = 700$, independently of policy type, phone brand, or model. 
Each policy covers exactly the period of the next 360 days after the underwriting day. 
For claims, we simulate the three incidents 
through a competing risk model with the constant hazards in Table \ref{tab:incidents}. 
All events are recorded daily and we identify a year with the grid $\{1/360,2/360,\dots,1\}$. After an incident has happened at time $T_0$, the reporting time is generated from the reporting delay hazard
\[
\alpha_0(t+T_0) = \frac{ t^{a-1} (1-t)^{b-1}}{\int_t ^1 s^{a-1} (1-s)^{b-1} \mathrm d s}, \ \ \ \ 0 < t < 1,
\]
for $a= 0.4$, $b = 10$. Hence, we assume a maximum reporting delay of one year. 
Denoting the reporting day by $T_1$, the payment day is then generated from the payment delay hazard 
\[
\alpha_1(t+T_1) = \frac{  (t-d/m)^{a-1} (1-(t-d/m))^{b-1}}{\int_{(t-d/m)}^1 s^{a-1} (1-s)^{b-1} \mathrm d s}, \ \ \ \ d < t < m+d,
\]
with $a=7$, $b=7$ and where $m=40/360$ and $d=10/360$. Hence, 
all claims are settled within 10 to 50 days. 
Note that both delays are independent of the incident or underwriting day $T_0$ and $T_1$ and, thus,  Assumptions [M3] and [CLM1] are satisfied. 
Moreover, reporting delay $\alpha_0$ and payment delay $\alpha_1$ are both independent of phone brand, model, type of policy, and type of incident. 
Last, we assume that the whole claim is settled in a single payment (Assumption [M2]) which is a random proportion of the phone price following a beta distribution with parameters given in Table \ref{tab:incidents}. 
\begin{table}[h!]
\centering
\begin{tabular}{llll}
  \hline
Incident  & Yearly hazard rate& $\alpha$ & $\beta$  \\ 
  \hline
Breakage & 0.15 & 2 & 5\\ 
Oxidation & 0.05 & 5 & 3\\
Theft & $ 0.05 \times \text{model} $&  5 & 0.5  \\
   \hline
\end{tabular}
\caption{Incidence hazard rates and parameters of the beta distribution, which determines the proportion of the phone price that is paid for a claim. }
\label{tab:incidents}
\end{table}

\begin{figure}[h!]
\centering
\subfigure[][Reserve estimates from September 2016 to May 2018. ]{
\includegraphics[width=0.45\textwidth]{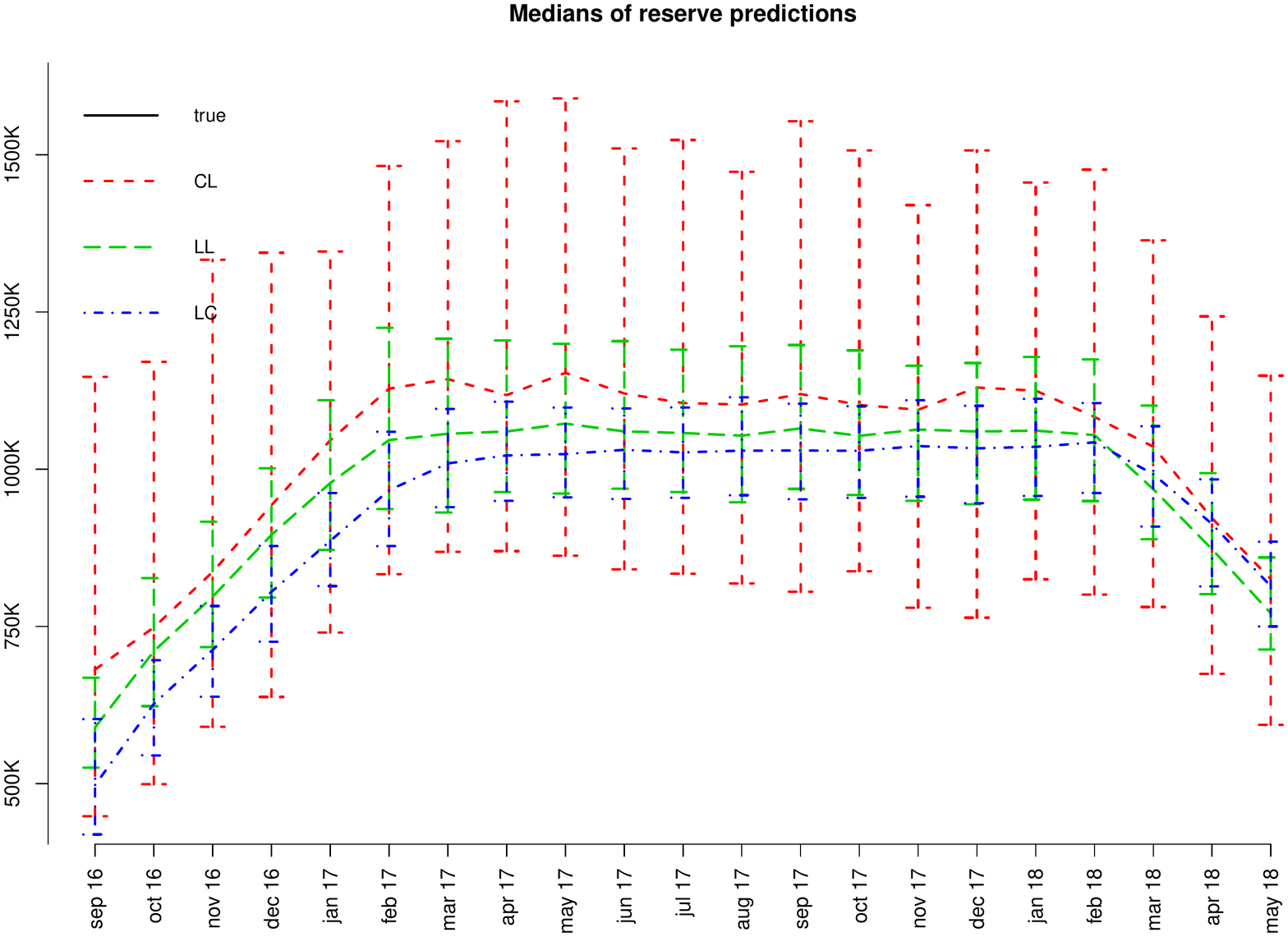}
}
\subfigure[][Mean squared error (MSE) in the reserve estimates from September 2016 to May 2018. ]{
\includegraphics[width=0.45\textwidth]{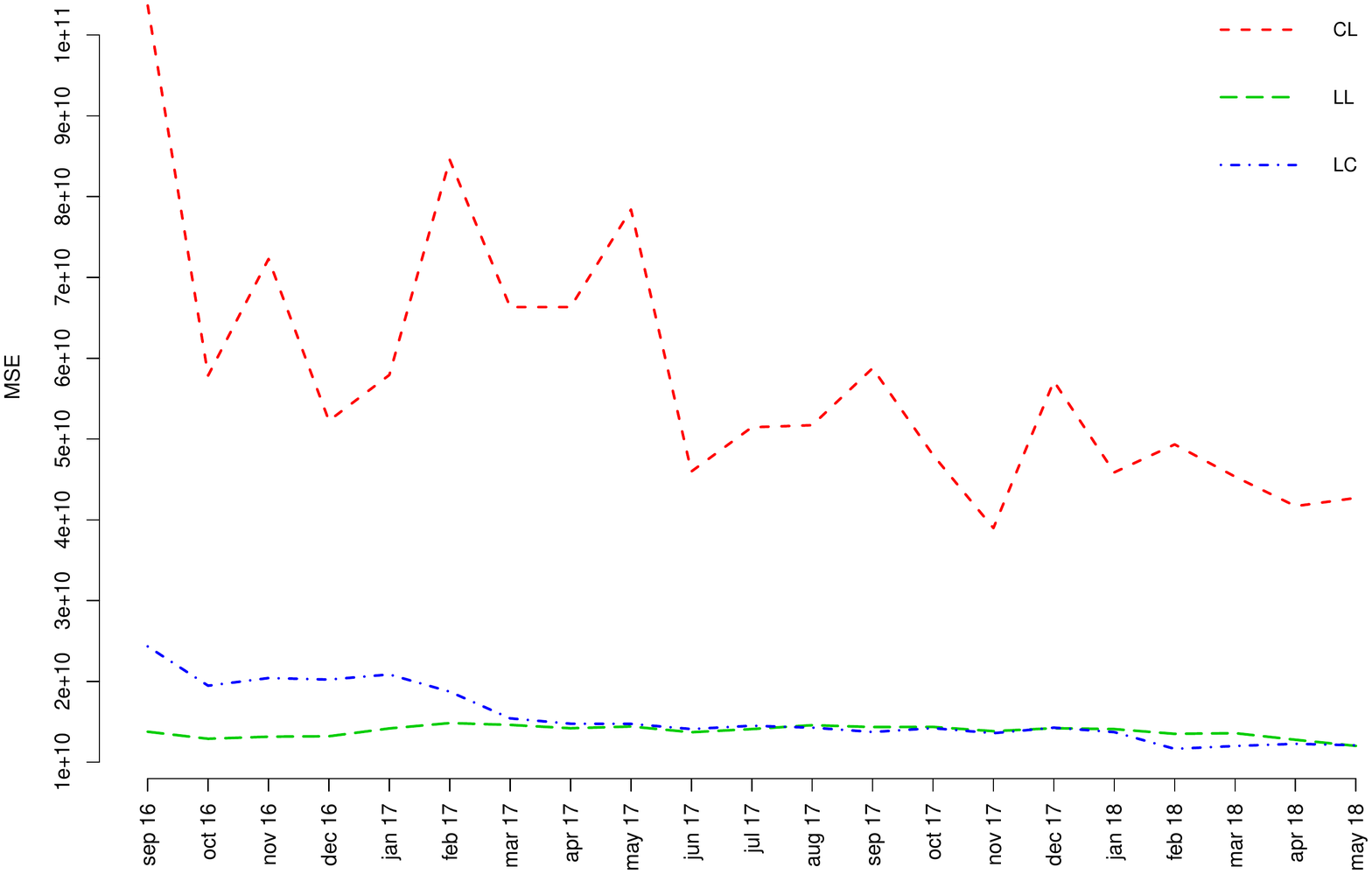}
}
\caption{Medians of estimates for reserves (a) and mean squared errors (b) over 200 simulation runs. Actual future payments (true), reserve estimate from the local linear (LL) density estimators, the local constant (LC) density estimators and chain-ladder (CL) with monthly aggregation. }
\label{fig:reserves:micro:data}
\end{figure}

We estimate the outstanding payments at each month from September 2016 to May 2018 with our new estimators and with monthly aggregated chain-ladder and compare it with the simulated future payments. The whole scenario is then repeated 200 times. 
The average reserves over all 200 simulation runs and their empirical 95\% confidence intervals are given in Figure \ref{fig:reserves:micro:data}a. It shows an increase in payments until February 2017, with new policies being underwritten every day. The payments stabilize afterwards in a balance between new policies and their claims and old policies expiring after 360 days. After December 2017, there is a decrease in payments since no new policies are underwritten anymore after 2017 and remaining policies expire. The medians of the actual outstanding future payments are taken over all 200 simulation runs and labeled as ``true" reserves. 
Moreover, the mean squared error 
\[
\MSE(\hat r_n, r_n)=200^{-1}\sum_{k=1}^{200} \left(\hat r^{[k]}_{n} - r^{[k]}_{n}\right)^2
\]
 for reserve estimate $\hat r_n^{[k]}$ and true reserve $r_n^{[k]}$ and simulation runs $k=1,\dots,200$ is given in Figure \ref{fig:reserves:micro:data}b. 
The reserve estimates from our local linear estimators $\widehat {\widetilde f}^{1,h}_{T}, \widehat {\widetilde f}^{1,h}_{U}$ have the lowest bias and variance. Whereas the local constant estimator suffers from bias, chain-ladder suffers heavily from variance (as found in \cite{Baudry:Robert:19}). 
The latter is due to the monthly aggregation for chain-ladder which, however, is necessary to derive the monthly cash-flow. Our proposed kernel smoothers use larger bandwidths, and thus reduce variance, while still providing  a monthly cash-flow. 
 
\begin{figure}[h!]
\centering
\subfigure[][Cost-weighted marginal distribution of delay. ]{
\includegraphics[width=0.45\textwidth]{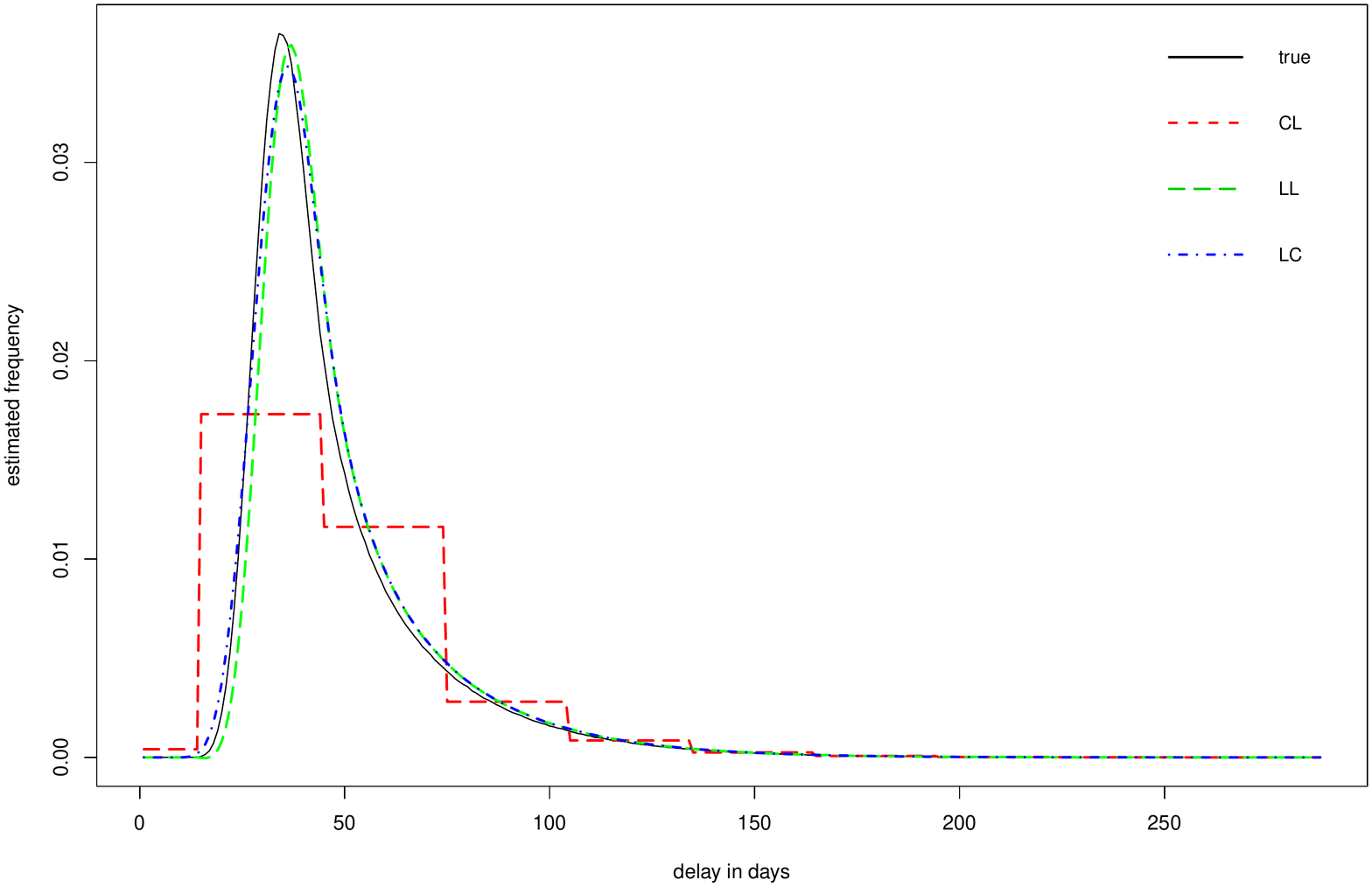}
}
\subfigure[][Cost-weighted marginal distribution of incident day and development factor type histogram. ]{
\includegraphics[width=0.45\textwidth]{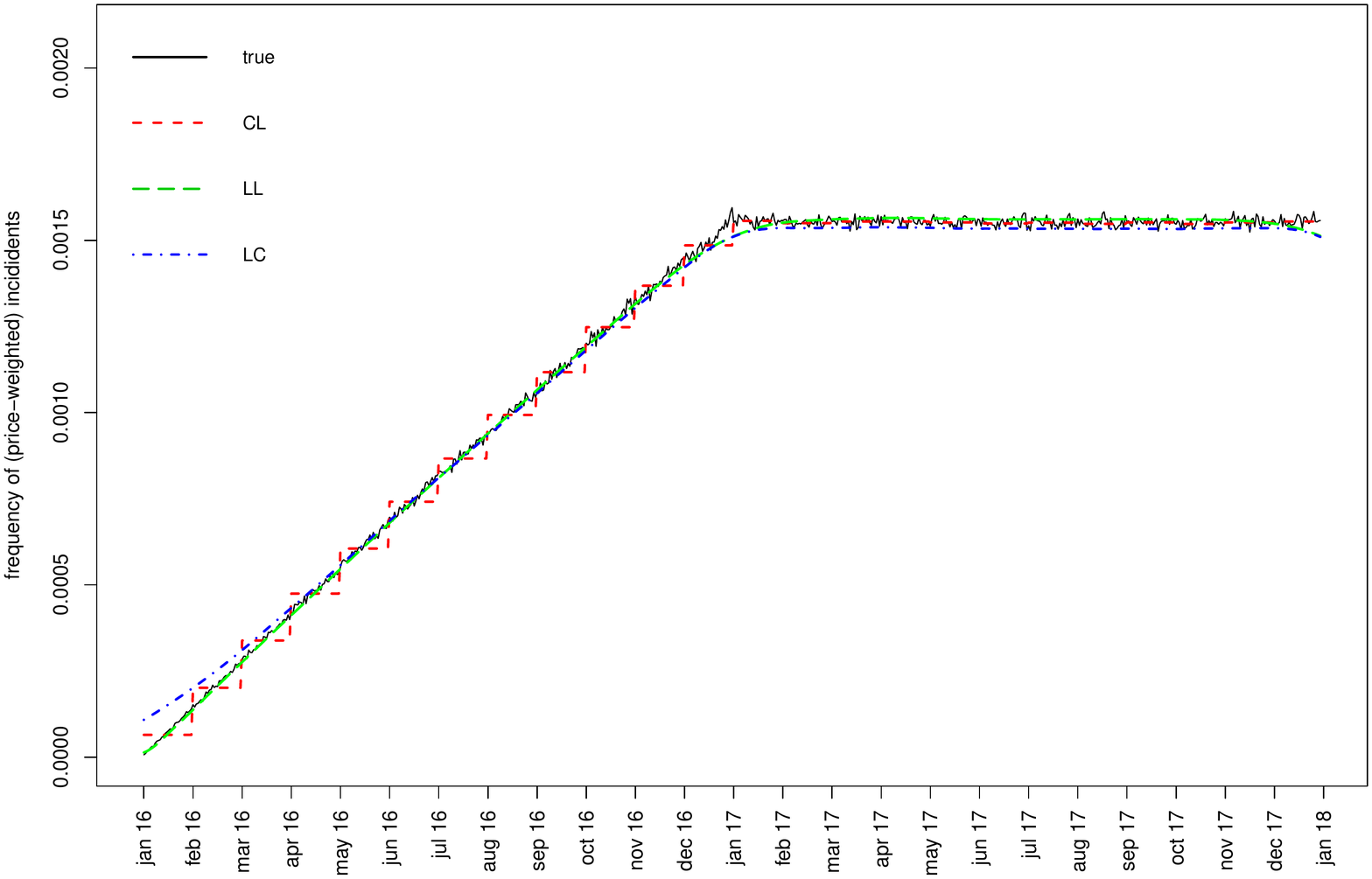}
}
\caption{Marginal distributions of the (cost-weighted) delay from incident day until payment day (a) and (cost-weighted) incident day (b), respectively, and monthly aggregated development factor histogram (CL), local linear estimators $\widehat {\widetilde f}^{1,h}_{T}, \widehat {\widetilde f}^{1,h}_{U}$ (LL), and local constant estimators $\widehat {\widetilde f}^{0,h}_{T}, \widehat {\widetilde f}^{0,h}_{U}$ (LC). }
\label{fig:distributions:micro:data}
\end{figure}

To compare this setting with the previous section, the marginal cost-weighted distributions of the delay from incident day until payment day and of the incident day are given in Figure \ref{fig:distributions:micro:data}a and b, respectively. The real distribution of the data is approximated by the average over the empirical distributions from each of the 200 simulation runs.  
Figure \ref{fig:distributions:micro:data}  illustrates how the development factors of the chain-ladder method lead to a histogram instead of a smooth kernel estimator as described in Section \ref{sec:CLM}.

\section*{Acknowledgments}
The authors would like to thank the editors and two anonymous referees for useful comments and suggestions which helped to improve this research article.

\appendix
\section{Proofs}
In the proofs below we will use the symbols $o_p(1)$ and $O_p(1)$ which are the probabilistic counterparts to the Landau symbols $o(1)$ and $O(1)$. A precise definition and explanation can be found in Appendix A of \cite{Pollard:12}. 
Further we will us the short-hand $\Delta N^R_i(t)=\lim_{h\downarrow 0} N^R_i\left\{(t+h)-\right\}- N^R_i(t-)$.
\subsection{Proof of Proposition \ref{prop11}}
\label{proof:prop1}
For the proof it suffices to show that
\begin{equation}\label{proof:prop1:1}
    \sup_{t\in[0,\mathcal{T}]}\Bigg|\frac{\sum_j Z_j Y^R_j(t)}{ \sum_j Y^R_j(t)}\Bigg| \xrightarrow{P}  \sup_{t\in[0,\mathcal{T}]}|E[Z_1 | Y^R_1(t)=1]|
\end{equation}
and 
\begin{equation}\label{proof:prop1:2}
    \frac{E\left[Z_1 | \ \Delta N^R_1(t)=1\right]} { E\left[Z_1 | \ Y^R_1(t)=1\right]} = \frac{E\left[Z | \ T^R=t \right]} { E\left[Z | \ T^R > t \right]},
\end{equation}
since $N_i$, $Y_i$, $Z_i$, $i=1,\dots,n$, are $iid$.

For the convergence of $(\ref{proof:prop1:1})$ we note that $\sup_{t\in[0,\mathcal{T}]}|n^{-1/2}\sum^n_{i=1}Y_i^R(t) - E[Y_i^R(t)]|= o_p(\log(n))$ and $\sup_{t\in[0,\mathcal{T}]}|n^{-1/2}\sum^n_{i=1}Y_i^R(t) Z_i - E[Y_i^R(t) Z_i]|= o_p(\log(n))$. Both statements follow from a strengthened Glivenko-Cantelli Theorem, since we have $Y^R_i(t)= I(U_i<t<T_i^R) = I(U_i<t) + I(T_i < \mathcal{T}-t) -1 $; see \cite[Chapter 19.1]{VanderVaart:00} for more details.
Next we argue that (\ref{proof:prop1:2})  is equivalent to [CLM2]. We  note that
\begin{align*}
\frac{E\left[Z_i | \ \Delta N^R_i(t)=1\right]} { E\left[Z_i | \ Y^R_i(t)=1\right]}&=\frac{E\left[Z| \ T^R=t, U \leq t \right]} { E\left[Z | \ T^R > t, U \leq t \right]} \\
&= \frac{\int_0^\infty \int_0^t zg(z,t,u) \mathrm du \mathrm dz  \int_t^{\mathcal T}\int_0^\infty \int_0^t g(z,s,u)\mathrm du \mathrm dz \mathrm ds }
{\int_t^{\mathcal T} \int_0^\infty \int_0^t z g(z,s,u) \mathrm du \mathrm dz \mathrm ds \int_0^\infty \int_0^t g(z,t,u) \mathrm du \mathrm dz.}
\end{align*}
Now, since $T^R$ and $U$ are independent,  we get
\begin{equation*}
\frac{\int_t^{\mathcal T}\int_0^\infty \int_0^t g(z,s,u)\mathrm du \mathrm dz \mathrm ds}{\int_0^\infty \int_0^t g(z,t,u) \mathrm du \mathrm dz} =
\frac{\int_t^{\mathcal T}\int_0^\infty \int_0^{\mathcal U} g(z,s,u)\mathrm du \mathrm dz \mathrm ds}{\int_0^\infty \int_0^{\mathcal U} g(z,t,u) \mathrm du \mathrm dz}=\alpha^R(t)^{-1}.
\end{equation*}
Hence equation (\ref{proof:prop1:2}) is equivalent to
\[
 \frac{\int_0^t \int_0^\infty zg(z,t,u) \mathrm dz \mathrm du }
{\int_t^1 \int_0^t \int_0^\infty z g(z,s,u) \mathrm dz \mathrm du \mathrm ds }=\frac{\int_0^1 \int_0^\infty zg(z,t,u) \mathrm dz \mathrm du }
{\int_t^1 \int_0^1 \int_0^\infty z g(z,s,u) \mathrm dz \mathrm du \mathrm ds}.
\]
With continuity arguments this holds if and only if  $\int_0^\infty z g(z,s,u) \mathrm dz$ is multiplicatively separable in $s$ and $u$, i.e, [CLM2] holds.

\subsection{Estimation of the weighted survival function} \label{appendix:survival}
We begin by investigating the asymptotic behaviour of 
\[
\widehat{\widetilde A}{}^{R,*}(t) = \sum_{i=1}^n \int_0^t \left\{\sum_{j\neq i}Z_jY_j^R(s)\right\}^{-1} \mathrm{d}\widetilde N^R_i(s)
\]
instead of $\widehat{\widetilde A}{}^R(t)$. Later with Lemma \ref{lem:substituteA}, we show that the difference between the two terms is uniformly of stochastic order $n^{-3/2}$ and hence negligible. 
One can hence carry this result over to a result for the actual estimators $\widehat{\widetilde S}{}^R$ and $\widehat {\widetilde f}^{0,h}$ or $\widehat {\widetilde f}^{1,h}$, respectively. 

To start with, we analyze the process  $\widehat{\widetilde A}{}^{R,*}_i(t)=\int_0^t\{\sum_{j\neq 1} Z_jY^R_j(s)\} \mathrm d \widetilde N^R_i(s)$, where the integral can be understood pathwise in Lebesgue-Stieltjes sense.
%

%
We start by deriving the compensator of $\widehat{\widetilde A}{}^{R,*}_i$:
\begin{align*}
&\lim_{h \downarrow 0} h^{-1}E\left[\widehat{\widetilde A}{}^{R,*}_i\left\{(t+h)-\right\}-\widehat{\widetilde A}{}^{R,*}_i(t-)|\ \mathcal F^R_{it-}\right] \\
=&\lim_{h \downarrow 0} h^{-1}E\left [\frac{Z_i}{\sum_{j\neq i} Z_jY_j^R(T_i^R)}I(T_i^R\in [t,t+h))| \mathcal F^R_{it-} \right]\\
=&\lim_{h \downarrow 0} h^{-1}E\left [\frac{Z_i}{\sum_{j\neq i} Z_jY_j^R(T_i^R)}|T_i^R\in [t,t+h) \right]E\left[N_i^R\left\{(t+h)-\right\}- N_i^R(t-)|\ \mathcal F^R_{it-}\right] \\
=& E[Z_i|\Delta N_i^R(t)=1] E\Bigg[ \frac{1}{\sum_{j\neq i} Z_jY_j^R(t)}\Bigg]  \alpha^R(t)Y_i^R(t)\\
=& \frac{E[Z_i|\Delta N_i^R(t)=1] }{(n-1) E[Z_iY_i^R(t)]}  \alpha^R(t)Y_i^R(t) + O(n^{-2})\\ 
= &\frac{ E[Z_i| \Delta N_i^R(t)=1]}{ (n-1)E[Z_i| Y_i^R(t)=1]\gamma(t)} \alpha^R(t)Y_i^R(t) + O(n^{-2}).
\end{align*}
Note that the error $O(n^{-2})$ comes from the Taylor expansion of $E[X^{-1}]$ at $E[X]$ for $X= \sum_{j\neq i} Z_jY_j^R(t)$. Since $Y_j^R(t)\leq 1$ for all $t$ and $Z_j \in L_1(\Omega)$, this error is also uniform of order $O(n^{-2})$.
Hence, 
\[
\widetilde \Lambda_i^R(t)=\frac {1} {(n-1)} \int_0^t \frac{E[Z_i| \Delta N_i^R(s)=1]}{ E[Z_i| Y_i^R(s)=1]\gamma(s)} \alpha^R(s)Y_i^R(s)\, \mathrm  d s , \quad (i=1,\dots,n),
\]
is asymptotically a compensator of the uniformly integrable submartingale $\widehat{\widetilde A}{}^{R,*}_i$.
We denote the resulting process by $\widetilde M^R_i=\widehat{\widetilde A}{}^{R,*}_i - \widetilde \Lambda_i^R$ which is, up to a lower order term of $O(n^{-2})$, a martingale.
Since $\widetilde M^R_i $ is cadlag with finite variation, the quadratic variation equals  the sum of square differences:
\[
[\widetilde M^R_i(t)]=\sum_{0<s\leq t} (\Delta \widetilde M^R_i(s))^2=\int_0^t \left\{\frac {Z_i}{\{\sum_{j\neq i} Z_jY_j^R(s)\}}\right\}^2 \mathrm d N_i^R(s).
\]
Note that we used $\Delta \widetilde M^R_i = \Delta \widehat{\widetilde A}{}^{R,*}_i$, since $\widetilde \Lambda_i^R$ is continuous. As $[\widetilde M^R_i(t)] = \left(\widehat{\widetilde A}{}^{R,*}_i\right)^2 $, by similar arguments as before we can calculate its compensator to derive the predictable variation process
\[
\langle \widetilde M^R_i(t)\rangle = \int_0^t \left\{\frac{E[Z_i| \Delta N_i^R(s)=1]}{ (n-1)E[Z_i| Y_i^R(s)=1]\gamma(s)}\right\}^2 \alpha(s)Y_i^R(s)\, \mathrm  d s .
\]
Proposition \ref{prop:2} is based on the following intermediate result.

\begin{lemma}\label{int:hazard:prop}
Under Assumptions [M1]--[M3], [CLM1]--[CLM2] and S1--S4, it holds that
\[
n^{1/2} \sum_{i=1}^n \widetilde M^R_i \rightarrow W(\sigma^2 ), \quad \sigma^2(t)=\int_0^t \left\{\frac{E[Z_i| \Delta N_i^R(s)=1]}{ E[Z_i| Y_i^R(s)=1]}\right\}^2 \alpha(s)\gamma^{-1}(s)\, \mathrm  d s,
\]
in distribution in Skorokhod topology sense, where $W$ is a zero mean Gaussian martingale with covariance, $\mathrm{Cov}\{W(s),W(t)\}=\sigma^2(s\wedge t)$.
\end{lemma}

\begin{proof}
This follows from a martingale central limit theorem in \cite{Rebolledo:80} as illustrated in \citet[p.~83]{Andersen:etal:93}.
For the assumptions to be satisfied, we verify that
\[
\langle n^{1/2} \sum_{i=1}^n \widetilde M^R_i(t)\rangle=n \sum_{i=1}^n \int_0^t   \left\{\frac{E[Z_i| \Delta N_i^R(s)=1]}{(n-1) E[Z_i| Y_i^R(s)=1]\gamma(s)}\right\}^2 \alpha(s)Y_i^R(s)\, \mathrm  d s\rightarrow \sigma^2(t),
\]
where we have used that $\langle \widetilde M^R_i, \widetilde M^R_j\rangle=0$ for $i\neq j$. 
For the Lindeberg condition we observe
\begin{align*}
&E\left[\sum_{s\leq t } \Delta \left( n^{1/2} \sum_{i=1}^n \widetilde M^R_i(t)\right) I{(\Delta ( n^{1/2} \sum_{i=1}^n \widetilde M^R_i(t))> \varepsilon)} \right] \\
=\,& n \sum_{i=1}^n E[\Delta \widehat{\widetilde A}{}^{R,*}_i(t_i) |  n^{1/2} \widehat{\widetilde A}{}^{R,*}_i(t_i) > \varepsilon] P(n^{1/2}\Delta \widehat A_i(t_i) > \varepsilon  ),
\end{align*}
where we used that the jumps $t_i$ happen at the same time with zero probability. The condition follows from the terms in the sum being $o(n^{-2})$, since $n^{1/2}Z_i(\sum_{j\neq i} Z_j Y_j^R(s))^{-1}\rightarrow 0$.
\end{proof}

\begin{corollary} \label{cor:survival:convergence}
Under Assumptions [M1]--[M3], [CLM1]--[CLM2] and S1--S4,  it holds that
\[
n^{1/2}\sup_t | \widehat{\widetilde S}{}^R(t) - \widetilde S^R(t)| = O_p(1).
\]
\end{corollary}
\begin{proof}
This follows from Lemma \ref{int:hazard:prop} with Lenglart's inequality and the functional delta method, since $\widehat{\widetilde S}{}^R$ and $\widetilde S^R$ are functionals of $\widehat {\widetilde A}{}^R$ and $\widetilde A^R$, respectively.
 
Indeed, it holds $\widetilde S^R(t) = \prod_{s\leq t} (1-\Delta \widetilde A^R(s))$ and $\widetilde A(t) = {(n-1)}{n}^{-1} \sum_{i=1}^n \hat \Lambda_i^R(t) + o_p(1)$. The conclusion from Lenglart's inequality (see \cite{Andersen:etal:93}) is
\[
P\left(n^{1/2} \sup_s | \widehat {\widetilde A}{}^R(s) - \widetilde A^R(s) | > \eta \right) \leq \frac{\delta}{\eta} + P\left( n^{1/2} \langle \sum_{i=1}^n \widetilde M^R_i (1) \rangle > \delta \right),
\]
for every $\delta>0$ and every $\eta>0$. Hence, the fact that $ n^{1/2} \langle \sum_{i=1}^n \widetilde M^R_i (1) \rangle \to \sigma^2(1) $ from the proof of Lemma \ref{int:hazard:prop} implies $n^{1/2} \sup_s | \widehat {\widetilde A}{}^R(s) - \widetilde A^R(s) | = O_p(1)$. Therefore, we get $n^{1/2}\sup_t | \widehat{\widetilde S}{}^R(t) - \widetilde S^R(t)| = O_p(1)$ \citep[p.~86]{Andersen:etal:93} and the conclusion follows.
\end{proof}

\subsection{Proof Proposition \ref{prop:2}}
\label{proof:prop2}
We first split the estimation error into a stable part and a martingale part via 
\[
\widehat {\widetilde f}^{R,0,h}_{T}-\widetilde f^R_T=B_0+V_0, \quad B_0=\widetilde f_T^{R,0,*}-\widetilde f^R_T, \quad V_0=\widehat {\widetilde f}_T^{R,0,h}-\widetilde f_T^{R,0,*}, 
\]
where
\[
 \widetilde f_T^{R,0,*}(t)=\frac{\sum_{i=1}^n\int_0^{\mathcal T}  K_{h}(t-s)\widehat{\widetilde S}{}^R(s)
E[Z_i| \ \Delta N_i^R(s)=1] Y_i^R(s)\alpha(s)\, \mathrm {d}s}{\sum_{i=1}^n\int_0^{\mathcal T} K_{h}(t-s) Z_iY_i^R(s)\, \mathrm {d}s}.
\]
We now discuss the asymptotics of $B_0$ and $V_0$ separately, starting with $V_0$. 
Defining
\[
\overline M^R_i= \int_0^{\mathcal T} \mathrm d \widetilde N_i^R(s) - \int_0^{\mathcal T} E\left[Z_i | \ \Delta N_i^R(s)=1 \right]\alpha(s) Y_i^R(s)\mathrm ds
\]
leads to 
\begin{align*}
 V_0(t)& =  \sum_{i=1}^n \int_0^{\mathcal T} \frac{K_h(t-s)\widehat{\widetilde S}{}^R(s)}{\sum_{i=1}^n\int K_h(t-s)Z_iY_i^R(s) \mathrm ds } \mathrm d \overline M^R_i(s) 
 \\
 &=   \sum_{i=1}^n \int_0^{\mathcal T} \frac{K_h(t-s)\widehat{\widetilde S}{}^R(s)}{n \int K_h(t-s) E[Z_1|\ Y_1^R(s)=1) ]\gamma(s)\mathrm ds } \mathrm d \overline M^R_i(s) + o_p(n^{-1/2}\log(n)),
\end{align*}
where the error is again uniformly bounded (see Proposition \ref{prop11}). Hence, since $\overline M^R_i$ is a martingale, 
$V_0$ is, up to lower order terms, a martingale as well. With similar arguments as before we get
\begin{equation*}
\langle \overline M^R_i (t)\rangle =  \int_0^t E\left[Z_i | \ \Delta N_i^R(s)=1 \right]^2\alpha(s) Y_i^R(s)\mathrm ds.
\end{equation*}
Similarly as in the proof of Lemma $\ref{int:hazard:prop}$, we use the martingale central limit theorem from \cite{Rebolledo:80} to show
\begin{equation} \label{V}
(nh)^{1/2} V_0 \rightarrow W(\sigma^2 ), \quad \sigma^2(t)=\{E\left[Z_i | \ T^R=t \right]/ E[Z_i]\}^2R( K) f(t) S^R(t) {\gamma(t)}^{-1},
\end{equation}
in distribution in Skorohod topology sense, where $W$ is a zero mean Gaussian martingale with covariance $\mathrm{Cov}\{W(s),W(t)\}=\sigma^2(s\wedge t)$. The assumptions are satisfied by similar arguments as in Lemma $\ref{int:hazard:prop}$. We only illustrate the derivation of $\sigma^2(t)$. It holds
\begin{align*}
&\langle (nh)^{1/2} \overline V_0(t)\rangle \\
=& nh \sum_{i=1}^n \int_0^{\mathcal T} \left\{\frac{K_h(t-s)\widehat{\widetilde S}{}^R(s)}{\sum_{i=1}^n \int_0^{\mathcal T} K_h(t-s)Z_iY_i^R(s) \mathrm ds }\right\}^2   E\left[Z_i | \ \Delta N_i^R(s)=1 \right]^2\alpha(s) Y_i^R(s)\mathrm ds \\
=&  \int_0^{\mathcal T} \left\{\frac{K(u)\widehat{\widetilde S}{}^R(t-uh)}{\int_0^{\mathcal T} K(u)\frac{1}{n}\sum_{i=1}^nZ_iY_i^R(t-uh) \mathrm du }\right\}^2 \\  & \quad \times E\left[Z_i | \ \Delta N_i^R(t-uh)=1 \right]^2\alpha(t-uh) \frac{1}{n} \sum_{i=1}^nY_i^R(t-uh)\mathrm du .
\end{align*}
With the uniform convergences of $\widehat{\widetilde S}{}^R$, $\sum_{i=1}^nY_i^R$ and $\sum_{i=1}^nZ_iY_i^R(t-uh)$, we conclude that
\begin{equation*}
\langle (nh)^{1/2} \overline V_0(t)\rangle \rightarrow \frac{R(K)\widetilde S^2(t)E[Z_i| \ \Delta N_i^R(s)=1]^2 \alpha(t)\gamma(t)}{\gamma(t)^2E[Z_i|  Y_i^R(t)=1]^2},
\end{equation*}
for $n\to\infty$, which coincides with $(\ref{V})$. 

We continue with the asymptotics for $B_0$.
After expanding $\widetilde{f}_T^{R,0,h}(t)$ and replacing $\widehat{\widetilde S}{}^R(s)$ by $\widetilde S^R(s) +  O_p(n^{-1/2})$, which we can do with 
Corollary \ref{cor:survival:convergence}, we have that
\[
B_0(t)= \frac{\sum_{i=1}^n\int_0^{\mathcal T}  K_{h}(t-s)Y_i^R(s)\{\widetilde f_T^R(s) E[Z_i| Y_i^R(s)=1]-Z_i \widetilde f (t)\}\mathrm ds }
{\sum_{i=1}^n\int_0^{\mathcal T} K_{h}(t-s) Z_iY_i^R(s)\mathrm {d}s}+ o(h^2).
\]
From the remark in A.1 we can further use that $n^{-1}\sum_{i=1}^n Z_i Y_i^R(s)$ converges uniformly to $E[Z_1| Y^R_1=1]\gamma(s)$ and it even holds that $\sup_{s\in [0,\mathcal{T}]} | n^{-1}\sum_{i=1}^n \{ Z_i Y_i^R(s)  -  E[Z_1 | Y^R_1(s)=1]\gamma(s) \} | =o_p(n^{-1/2}\log(n))$.  
Hence,
\[
B_0(t)= \frac{\int_0^{\mathcal T}  K_{h}(t-s)E[Z_1| Y^R_1(s)=1]\gamma(s)\{\widetilde f^R_T(s)-\widetilde f_T^R(t)\}\mathrm ds }
{\int_0^{\mathcal T} K_{h}(t-s) E[Z_1| Y^R_1(s)=1]\gamma(s) \mathrm {d}s}+ o(h^2)
\]
Note that an error term of order $O(n^{-2})$ is likewise of order $o(h^2)$ by Assumption S1. The proof  is concluded by two Taylor expansions in the numerator and one in the denominator and using that $K$ is a second order kernel. 

The implication for the estimator built from $\widehat{\widetilde A}{}^{R,*}$ instead of  $\widehat{\widetilde A}{}^{R}$ follows with Lemma \ref{lem:substituteA}.

\begin{lemma} \label{lem:substituteA}
It holds 
$n^{3/2} \sup_{s\in [0,\mathcal{T}]} | \widehat A(s) - \widehat A^*(s) | = O_p(1)$.
\end{lemma}
\begin{proof}
The proof follows from the fact that $Z_i \geq 0$ and $Y_i^R\leq 1$, $i=1,\dots, n$ together with an analogous argumentation as in the last proof.

First, because of the non-negativity of $Z_i$ and the boundedness of $Y_i^R$, it holds
\begin{align*}
\widehat{\widetilde A}{}^{R}(t) - \widehat{\widetilde A}{}^{R,*}(t) &= \sum_{i=1}^n \int_0^t \left[\frac{Z_i}{\sum_{j\neq i} Z_jY_j^R(s)}- \frac{Z_i}{\sum_{j=1}^n Z_jY_j^R(s)}\right] \mathrm d N_i^R(s) \\
&= \sum_{i=1}^n \int_0^t\frac{Z_i^2 Y_i^R(s)}{\{\sum_{j\neq i} Z_jY_j^R(s)\}\{ \sum_{j=1}^n Z_jY_j^R(s)\}} \mathrm d N_i^R(s) \\
&\leq \sum_{i=1}^n \int_0^t \frac{Z_i^2}{\{\sum_{j\neq i} Z_jY_j^R(s)\}^2} \mathrm d N_i^R(s) \\
&= \sum_{i=1}^n \left(\widehat{\widetilde A}{}^{R,*}_i\right)^2(t)
,
\end{align*}
for $t\in [0,1]$, with the notation $\widehat{\widetilde A}{}^{R,*}_i(t)$ from Section \ref{appendix:survival}. From here, a completely analogous argumentation with a martingale $\widetilde M^{R,*}_i(t) = \Big(\widehat{\widetilde A}{}^{R,*}_i\Big)^2 - \Lambda_i^{R,*}(t)$ for the compensator 
\[
\widetilde \Lambda_i^{R,*}(t)=\frac {1} {(n-1)^2} \int_0^t \left\{\frac{E[Z_1| \Delta N^R_1(s)=1]}{ E[Z_1| Y^R_1(s)=1]\gamma(s)}\right\}^2 \alpha(s)Y_i^R(s)\, \mathrm  d s , \quad (i=1,\dots,n)
\]
leads to the central limit theorem
\[
n^{3/2} \sum_{i=1}^n \widetilde M^R_i \rightarrow W(\sigma^2 ), \quad \sigma^2(t)=\int_0^t \left\{\frac{E[Z_1| \Delta N^R_1(s)=1]}{ E[Z_1| Y^R_1(s)=1]}\right\}^4 \alpha(s)\gamma^{-1}(s)\, \mathrm  d s.
\]
The same argument with Lenglart's inequality as in the proof of Corollary \ref{cor:survival:convergence} yields  the conclusion.
\end{proof}

\subsection{Proof of Proposition \ref{prop:3}} \label{proof:prop3}
We first introduce the notation
\[
\overline K^*(u)= \frac {\mu_2(K)-\mu_1(K)u}{\mu_2(K)-\left\{\mu_1(K)\right\}^2}K(u)
\]
for every kernel $K$ and remind of the notation
\[
\mu_j(K)=\int s^j K(s) \mathrm ds.
\]
Since 
\begin{equation}
\sup_{t\in [h,1-h]} |a_j(t)-h^j\mu_j(K)g(t)\gamma(t)|=o_p(1) \quad (j=1,2,3), \label{kernel:asymptotic}
\end{equation}
one can easily verify that  $n^{-1}\sum_i \overline K_{t,h}(t-s)\widetilde  Y_i^R(s)$  converges locally uniform almost surely to $\overline K^*_h(t-s)$,
where  $\overline K^*_h$ arises from $\overline K^*$ by replacing   $u$ and $K(u)$ with the local versions  $h^{-1}u, h^{-1}K(u/h)$ 
\citep{Nielsen:Tanggaard:01}. 
Furthermore, if $K$ is symmetric, then $\overline K^*(t)=K(t)$. \\

From equation \eqref{kernel:asymptotic}, Assumption S3 and Corollary \ref{cor:survival:convergence}, we conclude  that it is enough to consider the asymptotic behaviour of
\[ 
n^{-1} \sum_{i=1}^n \int_0^{\mathcal T} K_h(t-s)Z_i \widetilde S^R(s) \mathrm dN_i^R(s).
\]
Analogously to the local constant case, we split the estimation error into a stable and a martingale part
\[
B_1=\widetilde f_T^{R,1,*}-\widetilde f^R_T+ o_p(n^{-1/2}), \quad V_1=\widehat {\widetilde f}^{R,1,h}_{T} -\widetilde f_T^{R,1,*}+ o_p(n^{-1/2}), 
\]
where
\[
\widetilde f_T^{R,1,*}(t)=\int_0^{\mathcal T}  K_{h}(t-s) \widetilde f^R_T(s) \mathrm {d}s.
\]
The asymptotic limit of the bias part, $B_1$,  is now easily derived via a second order Taylor expansion.
The martingale part can be concluded with similar arguments as in Appendix \ref{appendix:survival}.

\subsection{Proof of Proposition \ref{prop:dv}} \label{proof:prop4}
The proof follows along the same lines as the proof of Proposition \ref{prop:2} above, just simpler,
with the choice 
\[
 \widetilde \alpha_H^{R,*}(t)=\frac{\sum_{i=1}^n\int_0^{\mathcal T}  I\{t\in[t_l,t_{l+1}\} 
E[Z_i| \ \Delta N_i^R(s)=1] Y_i^R(s)\alpha(s)\, \mathrm {d}s}{\sum_{i=1}^n I\{t\in[t_l,t_{l+1}\} Z_iY_i^R(s)\, \mathrm {d}s}.
\]

\bibliographystyle{agsm}  
\bibliography{references}
\end{document}

%% file: summary.res.tex
\begin{table}[h!]
\centering
\begin{tabular}{lrrrrrrr}
  \hline
  &  & LL &  & LC &  & CL &  \\ 
   &  & Median & Mean (s.d.) & Median & Mean (s.d.) & Median & Mean (s.d.) \\ 
   \hline
1 & 100 & 0.2937 & 0.3897 (0.4169) & 0.2821 & 0.4008 (0.4046) & 0.3525 & 0.5465 (0.4169) \\ 
   & 1000 & 0.0999 & 0.1198 (0.0911) & 0.0945 & 0.1170 (0.0952) & 0.1230 & 0.1459 (0.0911) \\ 
   & 10000 & 0.0376 & 0.0439 (0.0326) & 0.0353 & 0.0411 (0.0303) & 0.0425 & 0.0501 (0.0326) \\ 
   & 1e+05 & 0.0253 & 0.0259 (0.0155) & 0.0242 & 0.0251 (0.0146) & 0.0162 & 0.0186 (0.0155) \\ 
  2 & 100 & 0.3140 & 0.3892 (0.3650) & 0.2473 & 0.3743 (0.4307) & 0.3626 & 0.5370 (0.3650) \\ 
   & 1000 & 0.1094 & 0.1317 (0.1032) & 0.1164 & 0.1328 (0.0956) & 0.1266 & 0.1514 (0.1032) \\ 
   & 10000 & 0.0464 & 0.0541 (0.0380) & 0.0656 & 0.0694 (0.0434) & 0.0537 & 0.0583 (0.0380) \\ 
   & 1e+05 & 0.0302 & 0.0308 (0.0168) & 0.0429 & 0.0431 (0.0171) & 0.0256 & 0.0267 (0.0168) \\ 
  3 & 100 & 0.3350 & 0.3871 (0.3078) & 0.2773 & 0.3209 (0.2608) & 0.4443 & 0.6169 (0.3078) \\ 
   & 1000 & 0.1129 & 0.1309 (0.0979) & 0.1147 & 0.1317 (0.0943) & 0.1397 & 0.1581 (0.0979) \\ 
   & 10000 & 0.0510 & 0.0589 (0.0407) & 0.1056 & 0.1055 (0.0510) & 0.0962 & 0.0969 (0.0407) \\ 
   & 1e+05 & 0.0392 & 0.0393 (0.0192) & 0.1000 & 0.1001 (0.0189) & 0.0824 & 0.0821 (0.0192) \\ 
  4 & 100 & 0.3241 & 0.3552 (0.3022) & 0.4216 & 0.4199 (0.2195) & 0.4687 & 0.6987 (0.3022) \\ 
   & 1000 & 0.1198 & 0.1393 (0.1025) & 0.2368 & 0.2424 (0.1193) & 0.1979 & 0.2078 (0.1025) \\ 
   & 10000 & 0.0721 & 0.0776 (0.0497) & 0.1941 & 0.1939 (0.0537) & 0.1669 & 0.1664 (0.0497) \\ 
   & 1e+05 & 0.0475 & 0.0478 (0.0203) & 0.1650 & 0.1644 (0.0196) & 0.1445 & 0.1443 (0.0203) \\ 
  5 & 100 & 0.2871 & 4.189e-01 (0.8021) & 0.1877 & 2.831e-01 (0.3990) & 0.3198 & 1.663e+11 (0.8021) \\ 
   & 1000 & 0.1255 & 0.1399 (0.0952) & 0.1161 & 0.1286 (0.0869) & 0.1512 & 0.2437 (0.0952) \\ 
   & 10000 & 0.0490 & 0.0575 (0.0437) & 0.0660 & 0.0724 (0.0465) & 0.0763 & 0.0841 (0.0437) \\ 
   & 1e+05 & 0.0270 & 0.0293 (0.0195) & 0.0479 & 0.0479 (0.0210) & 0.0288 & 0.0319 (0.0195) \\ 
  6 & 100 & 0.3169 & 3.943e-01 (0.4260) & 0.2303 & 2.800e-01 (0.2365) & 0.3452 & 1.028e+10 (0.4260) \\ 
   & 1000 & 0.1214 & 0.1385 (0.1017) & 0.1228 & 0.1384 (0.0949) & 0.1502 & 0.2113 (0.1017) \\ 
   & 10000 & 0.0461 & 0.0539 (0.0408) & 0.0636 & 0.0694 (0.0454) & 0.0680 & 0.0758 (0.0408) \\ 
   & 1e+05 & 0.0295 & 0.0320 (0.0194) & 0.0522 & 0.0519 (0.0199) & 0.0301 & 0.0334 (0.0194) \\ 
  7 & 100 & 0.5853 & 7.398e-01 (1.0259) & 0.5016 & 5.832e-01 (0.6756) & 0.6462 & 6.066e+11 (1.0259) \\ 
   & 1000 & 0.2089 & 0.2954 (0.5423) & 0.2858 & 0.3258 (0.3315) & 0.3892 & 0.4983 (0.5423) \\ 
   & 10000 & 0.0981 & 0.1134 (0.1183) & 0.2109 & 0.2157 (0.1070) & 0.1802 & 0.1846 (0.1183) \\ 
   & 1e+05 & 0.0648 & 0.0675 (0.0414) & 0.1699 & 0.1702 (0.0568) & 0.1489 & 0.1478 (0.0414) \\ 
  8 & 100 & 0.4721 & 5.814e-01 (1.0504) & 0.4595 & 4.445e-01 (0.4085) & 0.6020 & 1.307e+11 (1.0504) \\ 
   & 1000 & 0.1661 & 0.2012 (0.1496) & 0.2717 & 0.2816 (0.1633) & 0.3565 & 0.4672 (0.1496) \\ 
   & 10000 & 0.1119 & 0.1217 (0.0822) & 0.2317 & 0.2303 (0.0908) & 0.2104 & 0.2059 (0.0822) \\ 
   & 1e+05 & 0.0934 & 0.0923 (0.0423) & 0.1962 & 0.1948 (0.0487) & 0.1819 & 0.1796 (0.0423) \\ 
   \hline
\end{tabular}
\caption{Median, mean and standard deviation of the squared relative errors $\err^2$ in the reserve estimate for 100, 1000, 10,000, and 100,000 observations. The statistics are taken over 1000 simulation runs. 
} 
\label{tab:reserve}
\end{table}

%% file: references.bib
@manual{R,
    title = {R: A Language and Environment for Statistical Computing},
    author = {{R Core Team}},
    organization = {R Foundation for Statistical Computing},
    address = {Vienna, Austria},
    year = {2018},
    url = {https://www.R-project.org/},
}

@article{Aalen:78,
   author =   {Aalen, O.~O.},
   title =    {Non-parametric inference for a family of counting processes},
   journal =     {The Annals of Statistics},
   year =     {1978 },
   volume =   {6},
   pages =    {701--726},
 }

@book{Andersen:etal:93,
	Address = {New York},
	Author = {Andersen, P. and Borgan, O. and  Gill, R. and Keiding, N.},
	Publisher = {Springer},
	Title = {Statistical Models Based on Counting Processes},
	Year = {1993},
}

@article{Antonio:Plat:14,
  title={Micro-level stochastic loss reserving for general insurance},
  author={Antonio, K. and Plat, R.},
  journal={Scandinavian Actuarial Journal},
  volume={2014},
  pages={649--669},
  year={2014},
}

@article{Arjas:89,
  title={The claims reserving problem in non-life insurance: Some structural ideas},
  author={Arjas, E.},
  journal={ASTIN Bulletin},
  volume={19},
  pages={139--152},
  year={1989}
}

@article{Austin:Betensky:14,
  title={Eliminating bias due to censoring in Kendall’s tau estimators for quasi-independence of truncation and failure},
  author={Austin, M.~D. and Betensky, R.~A.},
  journal={Computational Statistics \& Data Analysis},
  volume={73},
  pages={16--26},
  year={2014},
  publisher={Elsevier}
}

@article{Avanzi:etal:16,
  title={A micro-level claim count model with overdispersion and reporting delays},
  author={Avanzi, B. and Wong, B. and Yang, X.},
  journal={Insurance: Mathematics and Economics},
  volume={71},
  pages={1--14},
  year={2016}
}

@article{Badescu:etal:16,
  title={A marked {Cox} model for the number of {IBNR} claims: Theory},
  author={Badescu, A.~L. and Lin, X.~S.  and Tang, D.},
  journal={Insurance: Mathematics and Economics},
  volume={69},
  pages={29--37},
  year={2016}
}

@article{Baudry:Robert:19,
  author = {Baudry, M. and Robert, C.~Y.},
  title = {A machine learning approach for individual claims reserving in insurance}, 
  journal={Applied Stochastic Models in Business and Industry},
  publisher={Wiley Online Library},
  year = {2019}, 
  pages = {1--29},
}

@article{Bischofberger:etal:19a,
   author =   {Bischofberger, S.~M. and Hiabu, M. and Mammen, E.  and  Nielsen, J.~P.},
   title =    {A comparison of in-sample forecasting methods},
   journal =    {Computational Statistics \& Data Analysis},
   volume = {137},
   pages = {133--154},
   year = {2019},
 }

@article{Bowman:84,
   author =   {Bowman, A.~W.},
   title =    {An Alternative Method of Cross-Validation for the Smoothing of Density Estimates},
   journal =      {Biometrika},
   year =     {1984},
   volume =   {71},
   pages =    {353--360},
 }

@article{Cleveland:79,
  title={Robust locally weighted regression and smoothing scatterplots},
  author={Cleveland, W.~S.},
  journal={Journal of the American Statistical Association},
  volume={74},
  pages={829--836},
  year={1979}
}

@article{Crevecoeur:etal:19,
  title={Modeling the number of hidden events subject to observation delay},
  author={Crevecoeur, J. and Antonio, K. and Verbelen, R.},
  journal={European Journal of Operational Research},
  year={2019},
  publisher={Elsevier}
}

@article{England:Verrall:02,
   author =   {England, P.~D. and Verrall, R.~J.},
   title =    {Stochastic Claims Reserving In General Insurance},
   journal =      {British Actuarial Journal},
   year =     {2002},
   volume =   {8},
   pages =    {443--544},
 }

@book{Fan:Gijbels:96,
  title={Local polynomial modelling and its applications},
  author={Fan, J. and Gijbels, I.},
  year={1996},  
  publisher={Chapman and Hall},
  address={London}
}

@article{Gustafsson:etal:09,
  title={Local transformation kernel density estimation of loss distributions},
  author={Gustafsson, J. and Hagmann, M. and Nielsen, J.~P. and Scaillet, O.},
  journal={Journal of Business \& Economic Statistics},
  volume={27},
  pages={161--175},
  year={2009}
 }

@article{Hall:83,
   author =   {Hall, P.},
   title =    {Large sample optimality of least squares cross-validation in density estimation},
   journal =      {The Annals of Statistics},
   year =     {1983},
   volume =   {11},
   pages =    {1156--1174},
 }

@article{Hiabu:17,
   author =   {Hiabu, M.},
   title =    {On the relationship between classical chain ladder and granular reserving},
   journal =      {Scandinavian Actuarial Journal},
   year =     {2017},
   volume={2017},
   number={8},
   pages={708--729},
   publisher={Taylor \& Francis}
 }

@article{Hiabu:etal:16,
   author =   {Hiabu, M. and Mammen, E. and Mart{\'i}nez-Miranda, M.~D. and  Nielsen, J.~P.},
   title =    {In-sample forecasting with local linear survival densities},
   journal =      {Biometrika},
   year =     {2016},
   volume={103},
   pages={843--859},
 }

@article{Huang:etal:15,
  title={An individual loss reserving model with independent reporting and settlement},
  author={Huang, J. and Qiu, C. and Wu, X. and Zhou, X.},
  journal={Insurance: Mathematics and Economics},
  volume={64},
  pages={232--245},
  year={2015},
}

@article{Huang:etal:16,
  title={Asymptotic behaviors of stochastic reserving: Aggregate versus individual models},
  author={Huang, J. and  Wu,  X. and Zhou, X.},
  journal={European Journal of Operational Research},
  volume={249},
  pages={657--666},
  year={2016},
}

@article{Kremer:82,
  title={{IBNR-claims and the two-way model of ANOVA}},
  author={Kremer, E.},
  journal={Scandinavian Actuarial Journal},
  volume={1982},
  pages={47--55},
  year={1982},
}

@article{Kuang:etal:09,
   author =   {Kuang, D. and Nielsen, B. and Nielsen, J.~P.},
   title =    {Chain-ladder as maximum likelihood revisited},
   journal =      {Annals of Actuarial Science},
   year =     {2009},
   volume =   {4},
   pages =    {105--121},
 }

@article{Lee:etal:15,
   author =   {Lee, Y.~K. and  Mammen, E. and  Nielsen, J.~P. and Park, B.~U.},
   title =    {Asymptotics for In-Sample Density Forecasting},
   journal =      {The Annals of Statistics},
   year =     {2015},
   volume =   {43},
   pages =    {620--651},
 }

@article{Lee:etal:17,
   author =   {Lee, Y.~K. and  Mammen, E. and  Nielsen, J.~P. and Park, B.~U.},
   title =    {Operational time and in-sample density forecasting},
   journal =      {The Annals of Statistics},
   year =     {2017},
   volume =   {45},
   pages =    {1312--1341},
 }

@article{Mack:93,
  title={Distribution-free calculation of the standard error of chain ladder reserve estimates},
  author={Mack, T.},
  journal={Astin Bulletin},
  volume={23},
  pages={213--225},
  year={1993},
}

@book{Macaulay:31,
  title={The smoothing of time series},
  author={Macaulay, F.~R.},
  year={1931},
  publisher={National Bureau of Economic Research},
  address={New York}
}

@article{Martin:Betensky:05,
  title={Testing quasi-independence of failure and truncation times via conditional Kendall's tau},
  author={Martin, E.~C. and Betensky, R.~A.},
  journal={Journal of the American Statistical Association},
  volume={100},
  number={470},
  pages={484--492},
  year={2005},
  publisher={Taylor \& Francis}
}

@article{Martinez:etal:13,
   author =   {Mart{\'i}nez-Miranda, M.~D. and Nielsen, J.~P. and  Sperlich, S. and Verrall, R.~J.},
   title =    {Continuous Chain Ladder: Reformulating and generalising a classical insurance problem},
   journal =      {Expert Systems with Applications},
   year =     {2013},
   volume =   {40},
   pages =    {5588--5603},
 }

@article{Merz:etal:13,
  title={Dependence modelling in multivariate claims run-off triangles},
  author={Merz, M. and W{\"u}thrich, M.~V. and Hashorva, E.},
  journal={Annals of Actuarial Science},
  volume={7},
  pages={3--25},
  year={2013}
}

@article{Nielsen:Tanggaard:01,
   author =   {Nielsen, J.~P. and Tanggaard, C.},
   title =    {Boundary and bias correction in kernel hazard estimation},
   journal =      {Scandinavian Journal of Statistics},
   year =     {2001},
   volume =   {28},
   pages =    {675--698},
 }

@article{Nielsen:etal:09,
   author =    {Nielsen, J.~P. and Tanggaard, C. and Jones, M.~C.},
   title =    {Local linear density estimation for filtered survival data, with bias correction},
   journal =      {Statistics},
   year =     {2009},
   volume =   {43},
   number = {2},
   pages =    {167--186},
 }

@article{Norberg:93,
   author =   {Norberg, R.},
   title =    {Prediction of Outstanding Liabilities in Non-Life Insurance},
   journal =      {ASTIN Bulletin},
   year =     {1993},
   volume =   {23},
   pages =    {95--115},
 }

@book{Pollard:12,
  title={Convergence of stochastic processes},
  author={Pollard, D.},
  year={2012},
  publisher={Springer Science \& Business Media}
}

@article{Rebolledo:80,
  title={Central limit theorems for local martingales},
  author={Rebolledo, R.},
  journal={Zeitschrift für Wahrscheinlichkeitstheorie und verwandte Gebiete},
  volume={51},
  number={3},
  pages={269--286},
  year={1980},
}

@article{Renshaw:Verrall:98,
  title={A stochastic model underlying the chain-ladder technique},
  author={Renshaw, A.~E. and Verrall, R.~J.},
  journal={British Actuarial Journal},
  volume={4},
  pages={903--923},
  year={1998},
}

@article{Rudemo:82,
   author =   {Rudemo, M.},
   title =    {Empirical Choice of Histograms and Kernel Density Estimators},
   journal =      {Scandinavian Journal of Statistics},
   year =     {1982},
   volume =   {9},
   number = {2},
   pages =    {65--78},
 }

@article{Shi:etal:12,
  title={A Bayesian log-normal model for multivariate loss reserving},
  author={Shi, P.  and Basu, S. and Meyers, G.~G.},
  journal={North American Actuarial Journal},
  volume={16},
  pages={29--51},
  year={2012}
}

@article{Stone:77,
  title={Consistent nonparametric regression},
  author={Stone, C.~J.},
  journal= {The Annals of Statistics},
  pages={595--620},
  year={1977},
   volume =   {5},
}

@book{Taylor:86,
  title={Claims reserving in non-life insurance},
  author={Taylor, G.~C.},
  year={1986},
  publisher={Elsevier Science Ltd},
  address = {Amsterdam},
}

@book{VanderVaart:00,
  title={Asymptotic statistics},
  author={Van der Vaart, A.~W.},
  volume={3},
  year={2000},
  publisher={Cambridge University Press}
}

@article{Verrall:91,
  title={Chain ladder and maximum likelihood},
  author={Verrall, R.~J.},
  journal={Journal of the Institute of Actuaries},
  volume={118},
  pages={489--499},
  year={1991},
}

@book{Wand:Jones:94,
  title={Kernel Smoothing},
  author={Wand, M.~P. and Jones, M.~C.},
  series={Chapman \& Hall/{CRC} Monographs on Statistics \& Applied Probability},
  year={1994},
  publisher={Taylor \& Francis}
}

@article{Ware:DeMets:76,
   author =   {Ware, J.~H. and DeMets, D.~L.},
   title =    {Reanalysis of some baboon descent data},
   journal =      {Biometrics},
   year =     {1976},
   volume =   {32},
   number ={2},
   pages =    {459--463},
 }
